\documentclass[aps,twocolumn,pra,amsmath,amssymb,superscriptaddress,floats,nobibnotes,nofootinbib]{revtex4}

\usepackage{amsmath}	
\usepackage{gensymb}
\usepackage{hyperref}
\hypersetup{colorlinks=true}
\usepackage{upgreek}
\usepackage{graphics}
\usepackage{hyperref}
\usepackage{epsfig}
\usepackage{color}
\usepackage{bm}
\usepackage{float}
\usepackage{ulem} 
\usepackage{dsfont}                 
\usepackage{wasysym}
\usepackage{multirow}

\usepackage{blindtext}

\usepackage{accents}

\usepackage{graphicx}

\graphicspath{{./}{./plots/}}

\usepackage{url}
\usepackage{multirow}

\newcommand{\upperRomannumeral}[1]{\uppercase\expandafter{\romannumeral#1}}

\begin{document}

\title{Unusual excitations and double-peak specific heat\\in a bond-alternating spin-$1$ $K$-$\Gamma$ chain}

\author{Qiang Luo}
\affiliation{Department of Physics, University of Toronto, Toronto, Ontario M5S 1A7, Canada}
\author{Shijie Hu}
\email[]{shijiehu@csrc.ac.cn}
\affiliation{Beijing Computational Science Research Center, Beijing 100084, China}
\author{Hae-Young Kee}
\email[]{hykee@physics.utoronto.ca}
\affiliation{Department of Physics, University of Toronto, Toronto, Ontario M5S 1A7, Canada}
\affiliation{Canadian Institute for Advanced Research, Toronto, Ontario, M5G 1Z8, Canada}

\date{\today}

\begin{abstract}
  One-dimensional gapped phases that avoid any symmetry breaking have drawn enduring attention.
  In this paper, we study such phases in a bond-alternating spin-1 $K$-$\Gamma$ chain built of a Kitaev ($K$) interaction and an off-diagonal $\Gamma$ term.
  In the case of isotropic bond strength, a Haldane phase, which resembles the ground state of a spin-$1$ Heisenberg chain, is identified in a wide region.
  A gapped Kitaev phase situated at dominant ferromagnetic and antiferromagnetic Kitaev limits is also found.
  The Kitaev phase has extremely short-range spin correlations and is characterized by finite $\mathbb{Z}_2$-valued quantities on bonds.
  Its lowest entanglement spectrum is unique, in contrast to the Haldane phase, whose entanglement spectrum is doubly degenerate.
  In addition, the Kitaev phase shows a double-peak structure in the specific heat at two different temperatures.
  In the pure Kitaev limit, the two peaks are representative of the development of short-range spin correlation at $T_h \simeq 0.5680$
  and the freezing of $\mathbb{Z}_2$ quantities at $T_l \simeq 0.0562$, respectively.
  By considering bond anisotropy, regions of Haldane phase and Kitaev phase are enlarged,
  accompanied by the emergence of dimerized phases and three distinct magnetically ordered states.
\end{abstract}


\maketitle

\section{Introduction}

The Kitaev honeycomb model \cite{Kitaev2006}, consisting of bond-dependent Ising couplings of spin-$1/2$ degrees of freedom,
is a rare example which not only is exactly solvable but also hosts a quantum spin liquid (QSL) ground state
with fractionalized excitations, e.g., itinerant Majorana fermions and localized fluxes \cite{KnolleKCM2014}.
These excitations account for the double-peak specific heat anomaly at two different energy scales \cite{NasuUM2015}.
During the last decade, a large family of rare-earth magnets which could realize bond-directional interactions have garnered huge interest,
providing avenues for the exploration of exotic phases of matter and emergent phenomena
(for a review, see Refs.~\cite{RauLeeKee2016,TakagiTJ2019}).
According to the Jackeli-Khaliullin mechanism \cite{Jackeli2009},
it was suggested that the Kitaev interaction with $J_{\rm eff} = 1/2$ moment could be realized in the $4d/5d$-electron honeycomb compounds
by the interplay of spin-orbit coupling and electron correlations.
Recently, new scenarios for the Kitaev interaction in $f$-electron systems have also been proposed (see Ref.~[\onlinecite{MotomeSJSK2020}] and references therein).
This theoretical progress as well as relevant material realizations, paves the way for hunting Kitaev QSLs;
yet this is hindered by several essential non-Kitaev terms, such as Heisenberg coupling and symmetric off-diagonal $\Gamma$ interaction \cite{RanLeeKeePRL2014,LuoNPJ2021}.

Aligning with the efforts to study the ground state and thermodynamical properties in the spin-1/2 Kitaev honeycomb model
\cite{Kitaev2006,KnolleKCM2014,NasuUM2015,FengZX2007,BaskaranMS2007},
high-spin analogs have also generated much interest \cite{BaskaranSenSha2008,OitmaaKS2018,Rousochatzakis2018},
inciting the materialization of Kitaev interaction in magnetic compounds with $S > 1/2$ \cite{XuFengKa2020}.
Recently, it has been suggested that strong spin-orbit coupling between anion sites together with strong Hund's coupling in the $e_g$ orbital
might be a practical way to achieve the spin-1 Kitaev interaction in real materials \cite{StaPerKee2019}.
In spite of there being no exact solution for higher-spin counterparts,
local conserved $\mathbb{Z}_2$ quantities could still be constructed \cite{BaskaranSenSha2008},
and several interesting phenomena of the spin-$1/2$ model, including the double-peak specific heat \cite{KogaTN2018}
and field-induced intermediate gapless QSL \cite{ZhuWengSheng2020,HickeyTrebst2020,KhaitKim2021},
could also be retained at least for the spin-1 case.

Notwithstanding the bidimensionality of real materials, quantum spin chains also play vital roles in understanding peculiar quantum phenomena
in two dimensions as they promote strong quantum fluctuations \cite{AgrBrkNis2018,YangKG2020,YangJKG2020,YangSN2021,YouGam2020,SorenseCGK2021,LuoBAKG2021,MetaBrenig2021}.
Over the past few decades, quantum spin chains have attracted broad attention for their ability
to host unconventional quantum criticality \cite{ChepigaMila2019,JiangMot2019} and topological phases \cite{HaldanePRL1983,ZhaoHuZhang2015}.
The Haldane phase in the antiferromagnetic (AFM) spin-1 Heisenberg chain is a paramount example which falls beyond Landau's paradigm of symmetry breaking,
and is now recognized as a symmetry-protected topological (SPT) phase protected by time-reversal symmetry, bond-centered inversion symmetry,
and/or a dihedral group of $\pi$ rotations about cubic axes \cite{GuWen2009,PollmannSPT2012}.
Its ground state is unique under the periodic boundary condition (PBC),
while it is four-fold degenerate under the open boundary condition (OBC) as a result of two spin-$1/2$ edge states \cite{Kennedy1990}.
Nevertheless, a novel ``Kitaev" phase, the ground state of the spin-1 Kitaev spin chain \cite{SenShankar2010,YouSunRen2020},
emerges as another interesting phase which is also gapped and has the same ground-state degeneracy pattern as that of the Haldane phase.

In this respect, a couple of attractive questions are naturally raised.
First of all, although the Kitaev phase owns all the three symmetries that protect the Haldane phase \cite{YouSunRen2020},
the origin of the edge states that contribute to the ground-state degeneracy is still unclear.
To understand whether it is a SPT phase or not, the entanglement spectrum is a useful quantity to clarify the puzzle.
Next, in the Kitaev phase, degeneracy of the lowest-lying excited state goes with the increase in system size,
exhibiting a large density of states just above the excitation gap.
This phenomenon has been demonstrated to trigger double-peak specific heat anomalies in several frustrated systems \cite{Syromyatnikov2004,KarStrMad2016}.
Hence, it is necessary to check the low-temperature behavior of the specific heat.
Finally, in the spin-$1/2$ analogy, the ferromagnetic (FM) Kitaev limit is known to be a multicritical point
as a confluence of several topological quantum phase transitions (QPTs) \cite{LuoBAKG2021}.
By contrast, a spin-1 FM Kitaev point should survive against competing interactions,
giving rise to a possible region of the Kitaev phase.

In this paper, we study the quantum phase diagram of a bond-alternating spin-1 $K$-$\Gamma$ chain
using the density-matrix renormalization group~(DMRG) method \cite{White1992,Peschel1999,Schollwock2005}.
For this model, it is composed of two bond-directional frustrated interactions,
allowing us to explore the rich phase diagram by tuning both the interaction intensity and the bond strength relatively.
Throughout the phase diagram, we identify the Kitaev phase and the Haldane phase in the vicinity of the Kitaev and $\Gamma$ limits, respectively.
The natures of phases are revealed by excitation gaps, spin-spin correlations,
the nonlocal string order parameter (SOP) \cite{denNijsRom1989}, and $\mathbb{Z}_2$ quantities on bonds.
Near the FM Kitaev limit, there is a first-order Kitaev--Haldane QPT at nonzero $\Gamma/|K|$,
while a magnetically ordered state intervenes in the AFM Kitaev region.
We also calculate thermodynamic quantities (e.g., specific heat and thermal entropy) of the Kitaev phase
using the transfer-matrix renormalization group~(TMRG) method \cite{BurXiangGeh1996,WangXiang1997}.

The remainder of the paper is organized as follows.
In Sec.~\ref{SEC:Model} we introduce the theoretical model and present the quantum phase diagram.
In Sec.~\ref{SEC:IsoKG} we study the excitations of the Haldane phase and Kitaev phase in the isotropic $K$-$\Gamma$ chain.
Effects of anisotropic bond strength are studied in Sec.~\ref{SEC:AnisoKG},
with an emphasis on the Haldane--dimer transition and three magnetically ordered states.
In Sec.~\ref{SEC:CvEE}, we confirm the double-peak specific heat in the Kitaev phase, and explain the physical origins of the two peaks.
Finally, a brief conclusion is presented in Sec.~\ref{SEC:CONC}.
Further information about the specific heat in the spin-$1/2$ and spin-1 Kitaev chains is given in Appendices A and B.

\section{Model and Method}\label{SEC:Model}
\begin{figure}[!ht]
\centering
\includegraphics[width=0.99\columnwidth, clip]{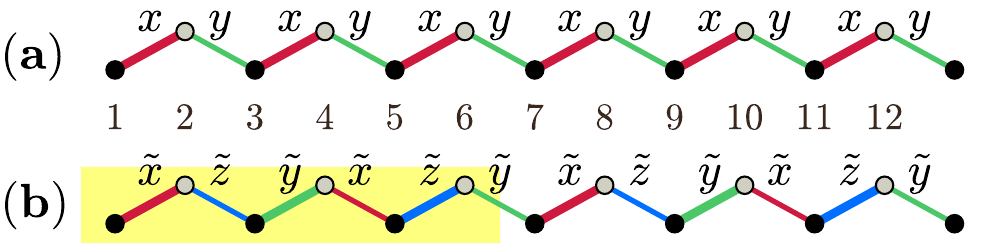}\\
\caption{(a) Sketch of the bond structure in the original form.
    Here, $x$~(red bonds) and $y$~(green bonds) stand for the $\gamma$-index and bond widths indicate their strengths relatively.
    (b) Pictorial bond structure of the rotated Hamiltonian. The shaded region represents a period of six sites in the $U_6$ rotation.}
    \label{FIG-Bond}
\end{figure}

The $K$-$\Gamma$ spin model is defined on the zigzag chain as illustrated in Fig.~\ref{FIG-Bond}(a),
where the spins sit on the edges of each bond.
The full Hamiltonian is composed of two analogical terms,
\begin{align}\label{J1J2KG-Ham}
\mathcal{H} = \sum_{l=1}^{L/2} g_x\mathcal{H}_{2l-1,2l}^{(x)}(\theta) + g_y\mathcal{H}_{2l,2l+1}^{(y)}(\theta),
\end{align}
where $L$ is the chain length and $g_x$~($g_y$) is the strength of the odd~(even) bond.
The exchange part contains the Kitaev ($K$) interaction and the off-diagonal $\Gamma$ interaction, which is given by
\begin{align}\label{KtvGam-Ham}
\mathcal{H}_{i,j}^{(\gamma)}(\theta) = K S_i^{\gamma}S_j^{\gamma} + \Gamma (S_i^{\alpha}S_j^{\beta}+S_i^{\beta}S_j^{\alpha}).
\end{align}
Here, $\gamma$ could be either $x$ or $y$ and it refers to the type of bond that connects spins $i$ and $j$, see Fig.~\ref{FIG-Bond}(a).
The triad of $\{\alpha, \beta, \gamma\}$ is $\{y, z, x\}$ on the $x$ bond and $\{z, x, y\}$ on the $y$ bond, respectively.
Following a $U_6$ rotation with a period of six sites,
all the cross terms in Eq.~\eqref{KtvGam-Ham} are eliminated, leading to the following form \cite{YangKG2020}
\begin{align}\label{U1XYZ-Ham}
\tilde{\mathcal{H}}_{i,j}^{(\gamma)}(\theta) =
-K \tilde{S}_i^{\gamma} \tilde{S}_j^{\gamma} - \Gamma (\tilde{S}_i^{\alpha}\tilde{S}_j^{\alpha}+\tilde{S}_i^{\beta}\tilde{S}_j^{\beta})
\end{align}
where the bonds $\gamma$ = $\tilde{x}$, $\tilde{y}$, and $\tilde{z}$ circularly, as depicted in Fig.~\ref{FIG-Bond}(b).
In light of Eq.~\eqref{U1XYZ-Ham}, it can instantly be found that $SU(2)$ symmetry recovers along two lines $K = \pm|\Gamma|$.

\begin{figure}[!ht]
\centering
\includegraphics[width=0.90\columnwidth, clip]{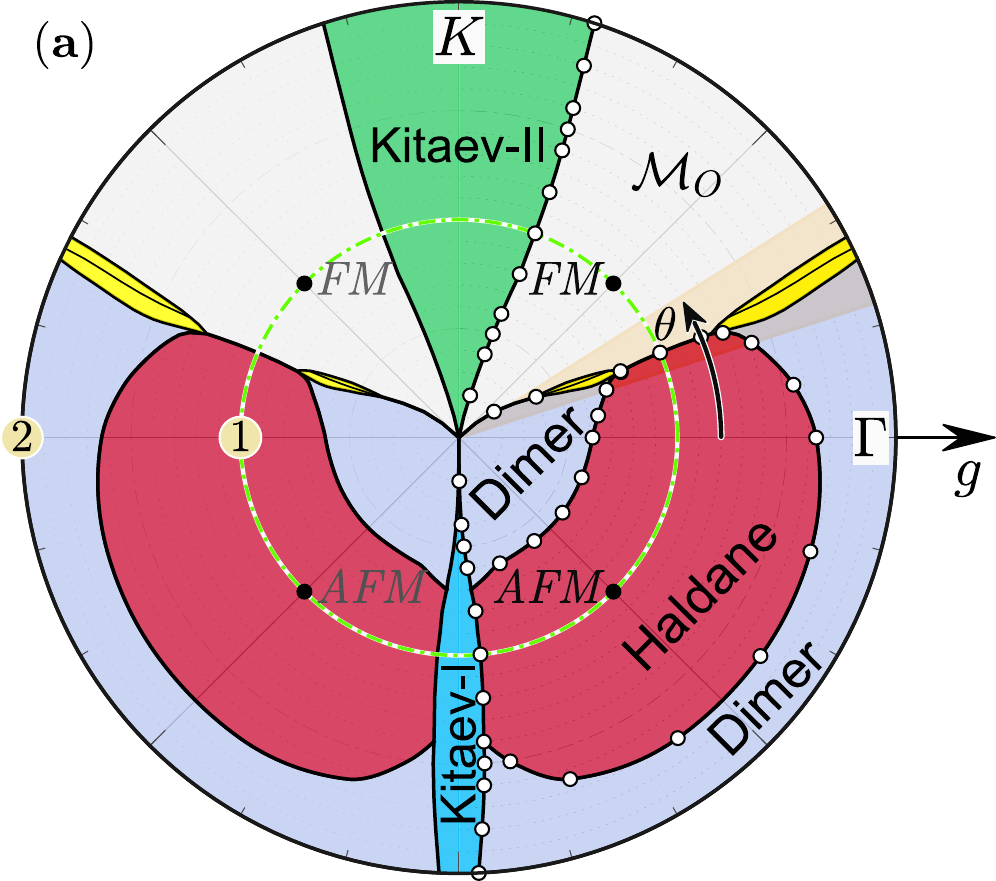}\\
\vspace{0.10cm}
\includegraphics[width=0.90\columnwidth, clip]{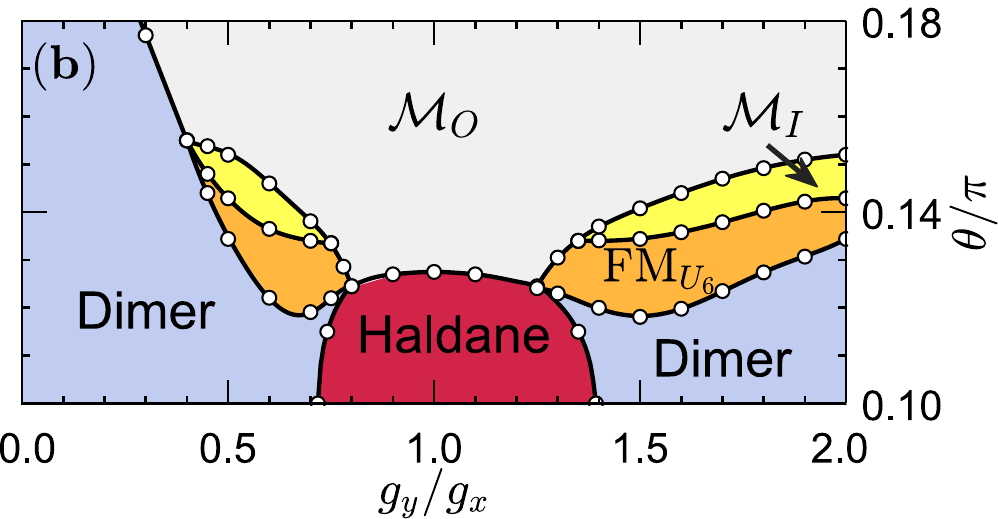}\\
\caption{(a) Quantum phase diagram of the bond-alternating spin-1 $K$-$\Gamma$ chain with $K=\sin\theta$ and $\Gamma=\cos\theta$.
    The green dotted line marked by $\textcircled{1}$ denotes the isotropic case of $g = 1$.
    Black solid circles at $\theta = \pi/4$ and $-\pi/4$ represent the hidden $SU(2)$ FM and AFM Heisenberg chains, respectively.
    In the right half circle, there is a Haldane phase, a gapped Kitaev phase whose two regions are distinguished as Kitaev-{\rm I} (FM Kitaev region) and Kitaev-{\rm II} (AFM Kitaev region) domains,
    two dimerized phases located at $g < 1$ and $g > 1$, respectively, and three magnetically ordered states termed FM$_{U_6}$, $\mathcal{M}_I$ and $\mathcal{M}_{O}$
    (see text for details).
    (b) Zoom-in region of $0.10\pi \leq \theta \leq 0.18\pi$.}
    \label{FIG-GSPD}
\end{figure}

As demonstrated in Ref.~[\onlinecite{LuoBAKG2021}],
the Hamiltonian in Eq.~\eqref{J1J2KG-Ham} possesses two peculiar properties concerning symmetries in the parameter space.
One is a self-dual relation, which implies that the eigenvalue $E$ of $\mathcal{H}$ satisfies $E(g) = gE(1/g)$
where $g \equiv g_y/g_x$ is the relative bond strength.
The other is a mirror symmetry with respect to the $\Gamma$ axis, i.e., $E(K, \Gamma) = E(K, -\Gamma)$.
These relations cause us to consider the phase diagram mainly in the right half circle of Fig.~\ref{FIG-GSPD}(a), where $\Gamma \geq 0$ and $g \in [0, 1]$.

The interactions are parameterized as $K=\sin\theta$ and $\Gamma=\cos\theta$ with $\theta \in (-\pi, \pi]$,
and the ground-state phase diagram is mapped out by the DMRG method \cite{White1992,Peschel1999,Schollwock2005}.
In the DMRG calculation, we mainly adopt the PBC to remove the edge effect,
while the OBC is also used to study edge excitations.
Given the structure of the unit cell in the $U_6$ rotated basis,
we choose the chain length $L$ to be strictly a multiple of six.
In addition, up to 2000 block states are kept so as to maintain a small truncation error of $\sim10^{-7}$ at most.

Figure~\ref{FIG-GSPD}(a) depicts the full phase diagram within the region $g \in [0, 2]$ and $\theta \in (-\pi, \pi]$
for the bond-alternating spin-1 $K$-$\Gamma$ chain.
In either the FM or AFM Kitaev limit, there is a gapped Kitaev phase,
whose regions shall be distinguished as the Kitaev-{\rm I} phase and the Kitaev-{\rm II} phase, respectively, for the sake of clarity.
The Kitaev phases in the two regions share the same ground-state and thermodynamic properties,
although the former is more fragile against $\Gamma$ interaction.
Remarkably, such an asymmetric stability of the Kitaev phases in the FM and AFM Kitaev limits also persists in the spin-$1/2$ analogy
and may be a general feature of the $K$-$\Gamma$ model.
For example, in the spin-$1/2$ $K$-$\Gamma$ chain, it is found that the FM Kitaev limit is merely a multicritical point
while there is a finite region near the AFM Kitaev side \cite{YangKG2020,YangSN2021,LuoBAKG2021}.
In the spin-$1/2$ honeycomb-lattice $K$-$\Gamma$ model, the region of the FM Kitaev QSL is considerably shrunken
when compared with its AFM counterpart \cite{RanLeeKeePRL2014}.
Of note is that the asymmetry is proposed to come from the interplay of two flux-pair hopping processes,
whose magnitudes depend crucially on the sign of the Kitaev interaction \cite{ZhangHZB2021}.
Going back to the spin-1 $K$-$\Gamma$ chain, by increasing $\Gamma$ interaction from the isotropic FM Kitaev limit,
the Kitaev-{\rm I} phase survives up to $|\Gamma|/K = -0.10(1)$,
followed by a Haldane phase which holds a nonlocal SOP, a finite excitation gap, and also two gapless edge modes.
The Haldane phase exists in a wide anisotropic region of $0.60 \lesssim g \lesssim 1.70$,
and two partially dimerized phases are then induced via continuous QPTs.
The difference between the dimerized phases lies in that there is a stronger $x$ ($y$)-type bond in the inner (outer) circle of $g = 1$.
Oppositely, the Kitaev-{\rm II} phase near the AFM Kitaev limit is more robust and occupies a larger territory.
When $K$ and $|\Gamma|$ are comparable, a magnetically ordered state with eight-fold ground-state degeneracy is favored.
In addition, two narrow regions of extra magnetic orderings are induced at modest anisotropy [see Fig.~\ref{FIG-GSPD}(b)].

\section{Isotropic $K$-$\Gamma$ chain}\label{SEC:IsoKG}

\subsection{Characters of Haldane phase}

\begin{figure}[!ht]
\centering
\includegraphics[width=0.95\columnwidth, clip]{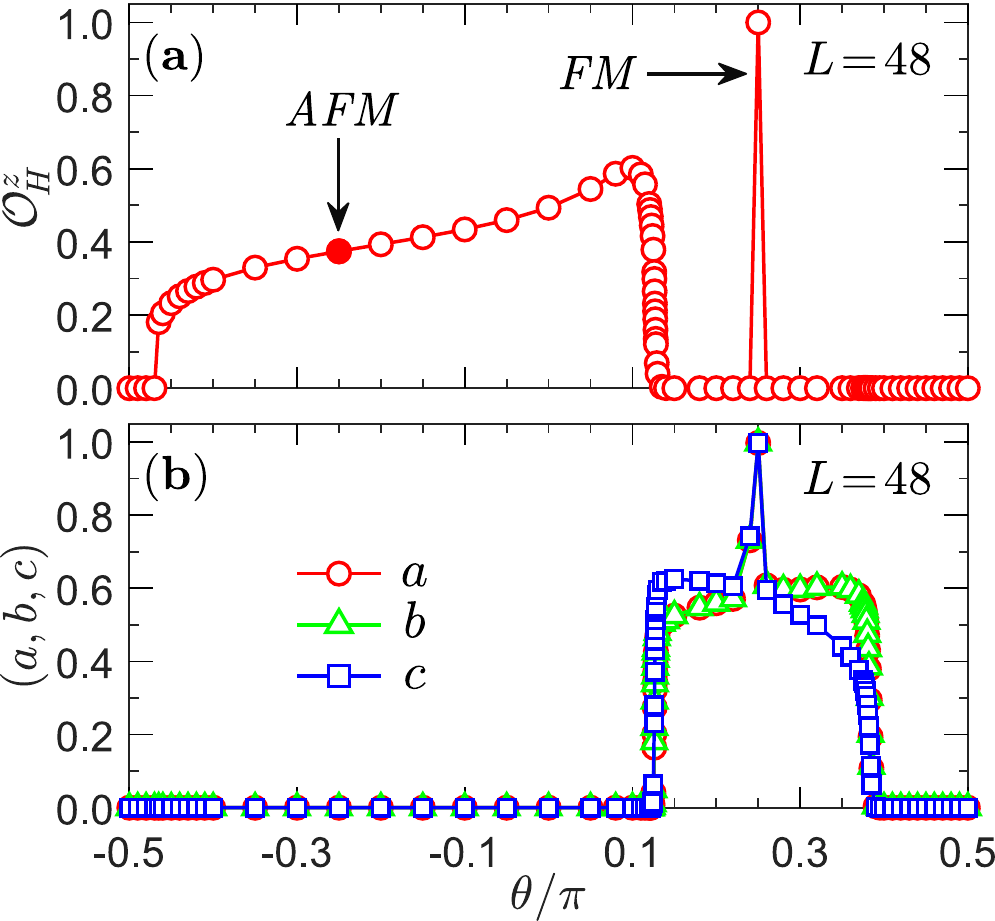}\\
\caption{(a) SOP $\mathcal{O}_H^z$ of the Haldane phase and (b) sublattice magnetization $(a, b, c)$ of the $\mathcal{M}_O$ phase
    in the isotropic $K$-$\Gamma$ chain with $g = 1$.
    The length $L$ of the chain is 48, and the results for $L = 96$ are extremely close (not shown) except for regions around transition points.
    }\label{FIG-KGSOPvsABC}
\end{figure}

First of all, let us consider the isotropic spin-1 $K$-$\Gamma$ chain with $g = 1$,
which is amenable to the revealment of crucial features of the entire phase diagram.
The $(K = -1, \Gamma = 1)$ point at $\theta = -\pi/4$ is a hidden $SU(2)$ AFM Heisenberg point whose ground state is the Haldane phase \cite{HaldanePRL1983}.
For this phase, it owns a (bulk) excitation gap $\Delta_e \simeq 0.410479$ \cite{WhiteHuse1993,EjimaFehske2015}
and possesses a nonlocal SOP defined as \cite{denNijsRom1989}
\begin{equation}\label{EQ:SOPHaldane}
\mathcal{O}_{H}^{z} = - \lim_{|q-p|\to\infty} \left\langle \tilde{S}_p^{z} \Big(\prod\limits_{p<r<q} e^{i\pi \tilde{S}_r^{z}}\Big) \tilde{S}_q^{z} \right\rangle
\end{equation}
where spins are situated in the $U_6$ rotated basis.
The gapped Haldane phase could survive against competing interactions with $\theta \neq -\pi/4$
since it is a SPT phase which usually undergoes a QPT with a closure of its excitation gap.
To estimate the region of the Haldane phase, we calculate the SOP $\mathcal{O}_{H}^{z}$ in the range of $\theta \in [-\pi/2, \pi/2]$,
as presented in Fig.~\ref{FIG-KGSOPvsABC}(a).
Except for the accidental FM point at $\theta = \pi/4$ where $\mathcal{O}_{H}^{z}$ is equal to the saturated value,
$\mathcal{O}_{H}^{z}$ is robust and suffers from a tiny finite-size effect in a wide region of $-0.4685(5) < \theta/\pi < 0.1270(5)$.
Here, the transition points and the corresponding error bars are estimated from the vanishing points of SOP.
These values are consistent with the independent estimate which shall be shown later.
With increasing $\theta$, $\mathcal{O}_{H}^{z}$ slowly grows deep in the Haldane phase.
Specially, the value of $\mathcal{O}_{H}^{z}$ is $0.4935(2)$ in the pure $\Gamma$ limit of $\theta = 0$,
which is larger than the value of 0.3743(1) for the AFM Heisenberg chain \cite{WhiteHuse1993}.
Near phase boundaries, $\mathcal{O}_{H}^{z}$ has a jump at $\theta_{t,1} = -0.4685(5)$ and varies smoothly around $\theta_{t,2} = 0.1270(5)$,
indicative of a first-order and a continuous QPT, respectively.

In the vicinity of the FM $SU(2)$ point, there is a magnetically ordered phase with an eight-fold ground-state degeneracy.
These degenerate states could be unified by the so-called $\eta$-notation \cite{RousochatzakisPerkins2017},
from which the spins within the six-site unit cell are
\begin{eqnarray}\label{EQ:M1Spin123}
\langle \tilde{\mathbf{S}}_1\rangle \!=\!
    \left(
    \begin{array}{c}
        \eta_x a \\
        \eta_y b \\
        \eta_z c
    \end{array}
    \right),
\langle \tilde{\mathbf{S}}_2\rangle \!=\!
    \left(
    \begin{array}{c}
        \eta_x a \\
        \eta_y c \\
        \eta_z b
    \end{array}
    \right),
\langle \tilde{\mathbf{S}}_3\rangle \!=\!
    \left(
    \begin{array}{c}
        \eta_x c \\
        \eta_y a \\
        \eta_z b
    \end{array}
    \right)
\end{eqnarray}
and
\begin{eqnarray}\label{EQ:M1Spin456}
\langle \tilde{\mathbf{S}}_4\rangle \!=\!
    \left(
    \begin{array}{c}
        \eta_x b \\
        \eta_y a \\
        \eta_z c
    \end{array}
    \right),
\langle \tilde{\mathbf{S}}_5\rangle \!=\!
    \left(
    \begin{array}{c}
        \eta_x b \\
        \eta_y c \\
        \eta_z a
    \end{array}
    \right),
\langle \tilde{\mathbf{S}}_6\rangle \!=\!
    \left(
    \begin{array}{c}
        \eta_x c \\
        \eta_y b \\
        \eta_z a
    \end{array}
    \right).
\end{eqnarray}
Here, $a$, $b$, $c$ $( \geq 0)$ satisfy the restriction $\sqrt{a^2+b^2+c^2} \leq S$ with $S = 1$,
while $\eta_{x}, \eta_y, \eta_z$ $( = \pm 1)$ are the Ising variables.
The three $\eta$'s are free to choose either 1 or $-1$ without altering the energy,
giving rise to the degenerate manifold.
Figure~\ref{FIG-KGSOPvsABC}(b) shows the values of $(a, b, c)$ in the same parameter region as Fig.~\ref{FIG-KGSOPvsABC}(a).
$a$ and $b$ are equal as a reminiscence of $g = 1$, and they compete with $c$ when changing the relative value of $K$ and $\Gamma$.
Since $c$ is always different from $a$ and $b$, the phase exhibits an out-of-plane spin structure
(cf. Eqs.~\eqref{EQ:M1Spin123} and \eqref{EQ:M1Spin456}) \cite{LuoBAKG2021} and thus is termed the $\mathcal{M}_O$ phase.
The $\mathcal{M}_O$ phase is found to exist in the parameter range of $\theta_{t,2} < \theta < \theta_{t,3}$ where $\theta_{t,3} = 0.3845(5)$.

The Haldane phase is known to carry a finite excitation gap, which is the key point of Haldane's conjecture \cite{HaldanePRL1983}.
We find that finite-size effects of the energy of the ground state and low-lying excited states are less pronounced in the PBC,
and the numerical result suggests that the Haldane phase indeed possesses a nonzero excitation gap in the whole region (not shown).
Taking the isotropic $\Gamma$-chain ($\theta = 0$) as an example,
our result indicates that the ground-state energy $e_g = -0.96226390(3)$ and the excitation gap $\Delta_e = 0.229135(7)$.
Noteworthily, the entanglement spectrum \cite{LiHaldane2008} is two-fold (four-fold) degenerate for the lowest-lying levels under the OBC (PBC),
which is a hallmark of the SPT phase \cite{PollmannTBO2010}.

\begin{figure}[!ht]
\centering
\includegraphics[width=0.95\columnwidth, clip]{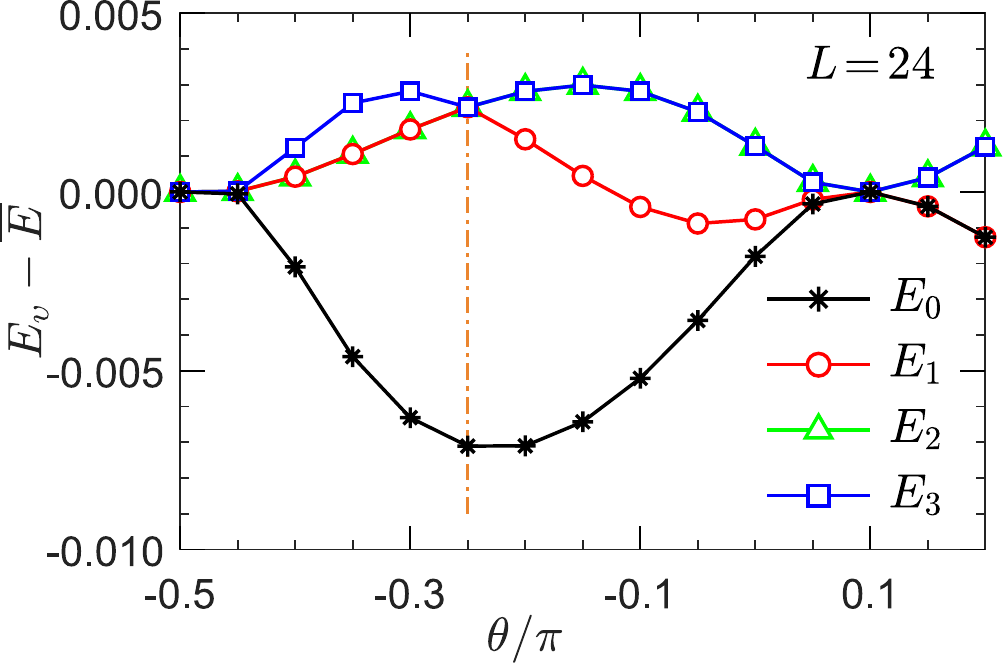}\\
\caption{Relative energy $\varepsilon_{\upsilon} = E_{\upsilon} - \bar{E}$ among the four quasi-degenerate ground states
    of the Haldane phase under an open chain with $L = 24$.
    The lowest energy level $E_0$ (black asterisk) is unique while higher ones are partially degenerate.
    }\label{FIG-HD4Deg}
\end{figure}

In addition, the Haldane phase also acquires two free edge spin-$1/2$s
which account for the four-fold degenerate ground state in the thermodynamical limit under the OBC \cite{WhiteHuse1993}.
In the isotropic spin-1 Heisenberg chain with an even number of sites,
the spin-$1/2$ edge modes are coupled with an effective AFM interaction.
Accordingly, the four-fold degeneracy is split into a single state and a Kennedy triplet state for finite-size systems \cite{Kennedy1990}.
However, this picture partially breaks down if $|K| \neq |\Gamma|$.
To explain this, we have calculated the first four energy levels $E_{\upsilon}$ ($\upsilon$ = 0--3) under an open chain with $L = 24$.
For the sake of comparison, we use the average energy $\bar{E} = (E_0+E_1+E_2+E_3)/4$ as a reference scale
and introduce a relative energy as $\varepsilon_{\upsilon} = E_{\upsilon} - \bar{E}$.
Figure~\ref{FIG-HD4Deg} presents $\varepsilon_{\upsilon}$ in the window of $\theta/\pi \in [-0.5, 0.2]$.
Throughout the Haldane phase, the lowest energy level (black asterisk) is always unique,
while higher ones are partially degenerate.
When $\theta < -\pi/4$ (i.e., $K < -|\Gamma|$), the quasi-degenerate states are of ``1+2+1" structure,
However, they are of ``1+1+2" structure when $\theta > -\pi/4$ (i.e., $K > -|\Gamma|$).
Higher levels recover as a Kennedy triplet state once $\theta = -\pi/4$ where the model is reduced to the spin-1 $SU(2)$ Heisenberg chain.

\begin{figure}[!ht]
\centering
\includegraphics[width=0.95\columnwidth, clip]{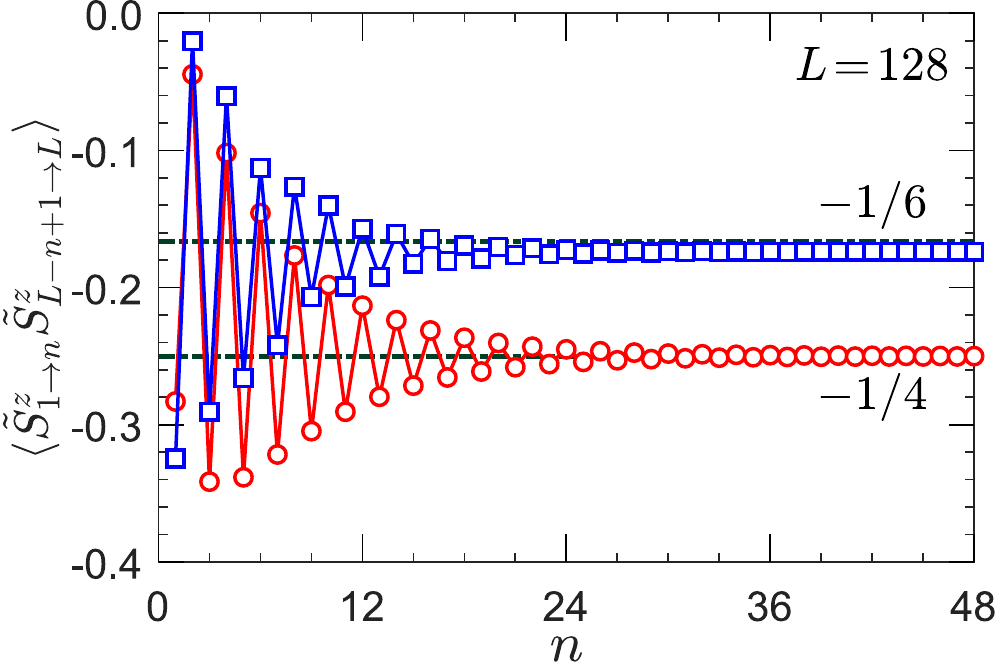}\\
\caption{The cumulant correlator $C_n^{z} = \langle \tilde{S}_{1 \to n}^z \tilde{S}_{L-n+1 \to L}^z\rangle$ as a function of segment $n$
    in the Haldane phase under an open chain with $L = 128$.
    $C_n^{z}$ approaches to $-1/4$ for the Heisenberg chain ($\theta = -\pi/4$, red circle) and around $-1/6$ for the $\Gamma$-chain ($\theta = 0$, blue square).
    }\label{FIG-HDEdgeMode}
\end{figure}

For the spin-1 Heisenberg chain, the existence of two edge spin-$1/2$s could be measured numerically
by calculating the on-site magnetization $\langle \tilde{S}_l^z\rangle$ in the $\tilde{S}_{\rm tot}^z = 1$ subspace,
which is found to decay exponentially towards the middle of the chain \cite{WhiteHuse1993,PolizziMilaSor1998}.
Notably, the edge states can also be probed by a couple of experimental methods (see Ref.~\cite{DelgadoBatFR2013} and references therein).
Unfortunately, detecting the edge modes is more intractable in our case for the lack of $U(1)$ symmetry.
In this regard, we adopt a cumulant correlation function
$C_n^{z} = \langle \tilde{S}_{1 \to n}^z \tilde{S}_{L-n+1 \to L}^z\rangle$,
where $\tilde{S}_{1 \to n}^z = \sum_{l=1}^n \tilde{S}_l^z$ and $\tilde{S}_{L-n+1 \to L}^z = \sum_{l=1}^n \tilde{S}_{L+1-l}^z$
are the accumulated magnetization on the left and right edges, respectively \cite{LiuNg2012}.
The $C_n^{z}$ represents the correlation between two edge spin-$1/2$s regardless of the singlet or triplet sector.
We calculate the cumulant correlation function $C_n^{z}$ on an open chain with 128 sites,
the size of which is far larger than the correlation length of the Haldane phase.
The behaviors of $C_n^{z}$ in the Heisenberg chain ($\theta = -\pi/4$) and $\Gamma$-chain ($\theta = 0$) are shown in Fig.~\ref{FIG-HDEdgeMode}.
While both curves have strong oscillations when $n$ is small,
they saturate to $-1/4$ and $-1/6$ or so, respectively, as $n \gtrsim 20$.
We note that, for the lack of Kitaev interaction, the correlator $C_n^{z}$ in the $\Gamma$-chain is thus only two-thirds of that in the Heisenberg chain.
Given the discrepancy between the saturated values of the correlator,
we confirm that there are two spin-$1/2$ edge modes which interact antiferromagnetically with each other as revealed by the minus sign of $C_n^{z}$.

\subsection{Unusual excitations of Kitaev phase}

We now turn to studying the nature of the Kitaev phase in the vicinity of two Kitaev limits.
In the spin-1 Kitaev spin chain,
the bond-directional Ising interactions allow for the existence of many $\mathbb{Z}_2$ quantities
that commute with the Hamiltonian.
To this end, it is natural to introduce an on-site operator
$\Sigma_l^{\alpha} = e^{i\pi S_l^{\alpha}}$ \cite{SenShankar2010},
which mutually commutes with each other and happens to be $\mathbb{I}-2(S_l^{\alpha})^2$ for the spin-1 case.
The bond-parity operators $\hat{W}_l$ on the odd/even bonds \cite{YouSunRen2020},
\begin{equation}\label{EQ:WbEvOd}
\hat{W}_{2l-1} = \Sigma_{2l-1}^{y}\Sigma_{2l}^{y},\quad
\hat{W}_{2l}   = \Sigma_{2l}^{x}\Sigma_{2l+1}^{x}
\end{equation}
commute with the Hamiltonian of the Kitaev spin chain with eigenvalues being $\pm1$ (which explains the $\mathbb{Z}_2$ nature).
In the Kitaev limit, the ground state is unique under the PBC and lies in the sector with all $\hat{W}_l = +1$ \cite{SenShankar2010}.
It is gapped with an extremely short correlation length $\xi \simeq 1$ since the spin-spin correlation beyond the nearest-neighbor bonds is extremely small.
The first excited state corresponds to flipping the eigenvalue of any of the bond parity operators to $-1$ while leaving the rest unchanged,
generating an $L$-fold degenerate first excited state \cite{YouSunRen2020}.
We find that the ground-state energy of the isotropic Kitaev chain is -0.603560592(3)\footnote{
The ground-state energy under OBC is $e_g = -0.6035606(2)$,
in accordance with the estimated value under PBC within the numerical precision.},
which matches perfectly with a previous estimate by exact diagonalization method \cite{SenShankar2010}.
Also of interest is that the energy is very close to that of the two-dimensional spin-1 Kitaev honeycomb model \cite{LeeKawKim2020}.

\begin{figure}[!ht]
\centering
\includegraphics[width=0.95\columnwidth, clip]{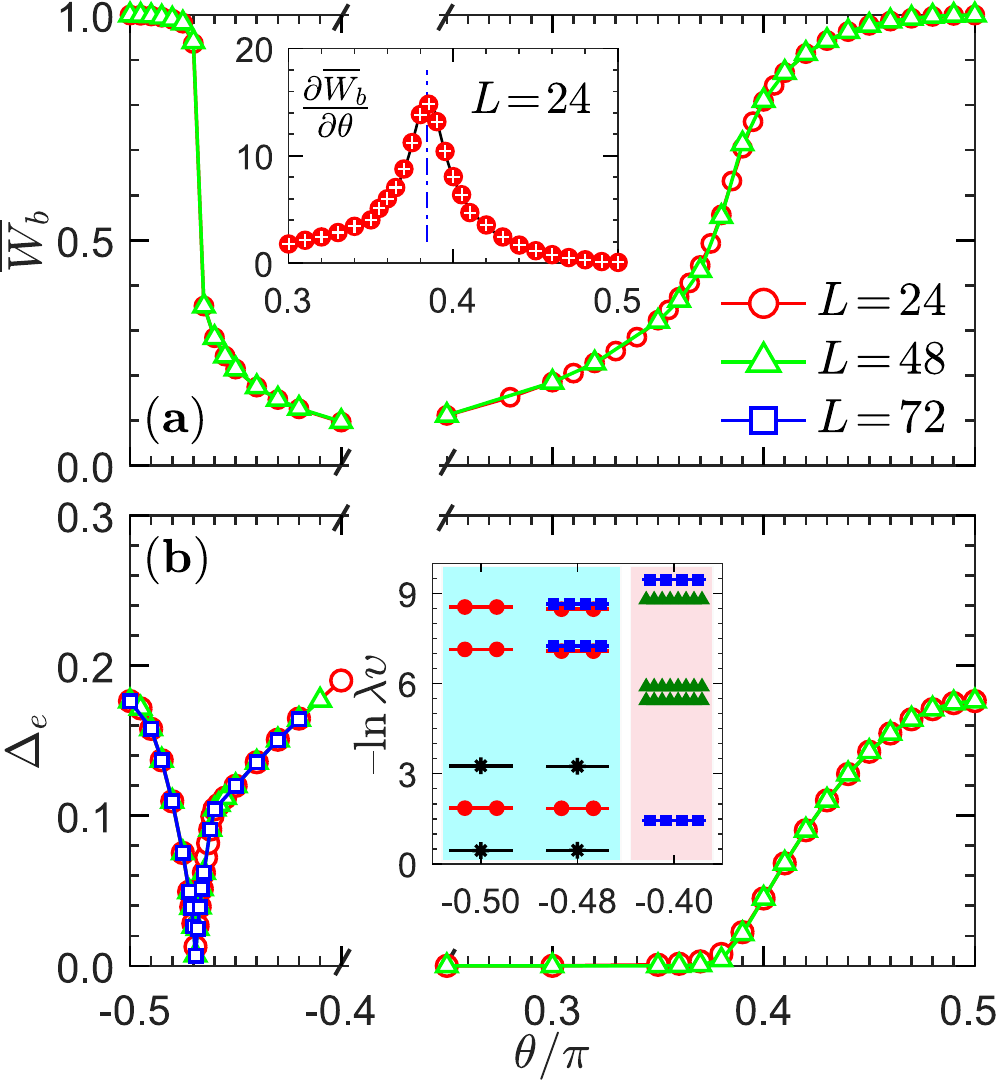}\\
\caption{(a) Bond density $\overline{W}_b$ and (b) excitation gap $\Delta_e$ of the Kitaev phase in the isotropic $K$-$\Gamma$ chain with $g = 1$.
    The length $L$ of the chain is 24 (red circle), 48 (green triangle), and 72 (blue square).
    The inset of (a) shows the first-order derivative of $\overline{W}_b$ with $L = 24$,
    while the inset of (b) presents the entanglement spectrum of the Kitaev phase ($\theta = -0.50\pi$ and $-0.49\pi$)
    and Haldane phase ($\theta = -0.40\pi$) with $L = 48$.}
    \label{FIG-KtvFluxGap}
\end{figure}

To capture the QPT driven by $\Gamma$ interaction, we define the averaged \textit{bond} density
\begin{equation}\label{EQ:WbAvg}
\overline{W}_{b} = \frac{1}{L}\sum_{l=1}^{L} \langle\hat{W}_{l}\rangle,
\end{equation}
which should still be $+1$ for the ground state in the Kitaev limit.
Figure~\ref{FIG-KtvFluxGap}(a) shows $\overline{W}_{b}$ as a function of $\theta$ for different length $L$ in both FM and AFM Kitaev limits.
In the left panel, $\overline{W}_{b}$ experiences a dramatic jump at $\theta_{t,1}/\pi = -0.4685(5)$ (i.e., $|\Gamma|/K \simeq -0.10$),
favoring the first-order QPT as revealed by the SOP of the neighboring Haldane phase (cf. Fig.~\ref{FIG-KGSOPvsABC}(a)).
However, $\overline{W}_{b}$ varies smoothly as $\theta$ decreases from $\pi/2$.
To locate the transition point, we take the first-order derivative of $\overline{W}_{b}$ with respect to $\theta$ (see inset).
It can be observed that there is a peak whose position is $\theta_{t,3}/\pi = 0.3845(5)$.
In Fig.~\ref{FIG-KtvFluxGap}(b) we show the excitation gap $\Delta_e = E_1 - E_g$,
defined as the energy difference between the ground state ($E_g$) and the first excited state ($E_1$), around the transition points.
In the range of $-0.50\pi \leq \theta \leq -0.40\pi$, both Kitaev-{\rm I} phase and Haldane phase are gapped,
and transition at $\theta_{t,1}$ is of first order due to the level crossing.
In addition, excitation gap of the Kitaev-{\rm II} phase at $\theta = 0.50\pi$ is equal to that of the Kitaev-{\rm I} phase at $\theta = -0.50\pi$,
with a value of $\Delta_e = 0.17634970(2)$.
Within the Kitaev-{\rm II} phase, $\Delta_e$ decreases gradually and is vanishingly small
when approaching the phase boundary of the Kitaev-{\rm II} to $\mathcal{M}_O$ transition.

So far, we have shown that the Kitaev--Haldane transition at $\theta_{t,1}/\pi = -0.4685(5)$ is a topological first-order transition
which can be recognized by the jumps in nonlocal SOP and $\mathbb{Z}_2$ quantities.
Both of the two are gapped with unique ground states.
The Haldane phase, in particular, is a preeminent example of SPT phase with short-range entanglement.
By contrast, the Kitaev phase is \textit{not} a SPT phase since its lowest entanglement spectrum is unique rather than doubly degenerate.
The entanglement spectrum is a quantity which encodes the spectral information of the subsystem $A$ with a reduced density matrix $\rho_A$.
It is defined as $-\ln\lambda_{\upsilon}$ where $\lambda_{\upsilon}$ is the eigenvalue of $\rho_A$ with $\sum_{\upsilon} \lambda_{\upsilon} = 1$ \cite{LiHaldane2008}.
The entanglement spectrum of the Haldane phase is universally known to be doubly degenerate;
it is two-fold degenerate under the OBC, while it is four-fold degenerate under the PBC \cite{PollmannTBO2010}.
As shown in the inset of Fig.~\ref{FIG-KtvFluxGap}(b),
the Haldane phase at $\theta = -0.40\pi$ indeed has a four-fold degeneracy entanglement spectrum under PBC.
However, spectra of the Kitaev phase at $\theta = -0.50\pi$ and $-0.48\pi$ are unique,
in contradiction to the character of a SPT phase.

\begin{figure}[!ht]
\centering
\includegraphics[width=0.95\columnwidth, clip]{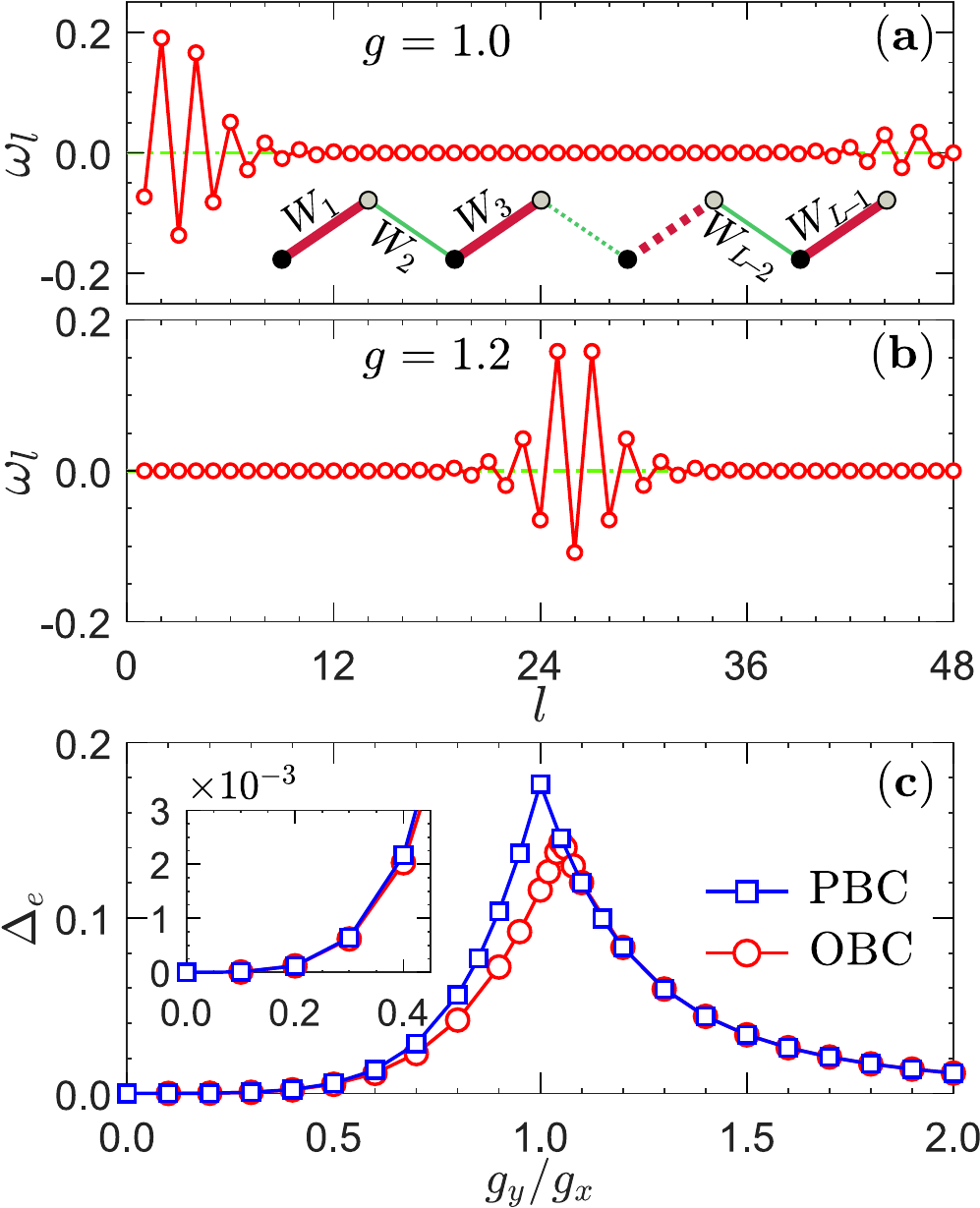}\\
\caption{Distribution of excitation energy $\omega_l$ in the whole system ($L = 48$) under the OBC with (a) $g = 1.0$ and (b) $g = 1.2$, respectively.
    The inset of (a) shows the positions of bond-dependent $\mathbb{Z}_2$ quantities $\hat{W}_{l}$ in a $L$-site chain.
    (c) Excitation gap $\Delta_e$ as a function of anisotropy $g$ under OBC (red circle) and PBC (blue square) in the thermodynamic limit ($L\to\infty$).}
    \label{FIG-KtvOBCGap}
\end{figure}

In the case of the OBC, the Haldane phase has a four-fold degenerate ground state owing to the two spin-$1/2$ edge states.
Interestingly, the ground state of the Kitaev phase under the OBC is also four-fold degenerate,
relating to the breakdown of two marginal $\mathbb{Z}_2$ quantities $\hat{W}_{1}$ and $\hat{W}_{L-1}$, see the inset of Fig.~\ref{FIG-KtvOBCGap}(a).
Here, both quantities are situated at odd bonds and their eigenvalues are fractional,
while the remaining $\mathbb{Z}_2$ quantities are $+1$.
Regarding values of $\hat{W}_{1}$ and $\hat{W}_{L-1}$ on the four-fold degenerate ground state,
they are somewhat random as a consequence of arbitrary superpositions,
and one of the computation on a 48-site open chain indicates that
$\langle\hat{W}_{1}\rangle \in \{0.6965, -0.8076, -0.8307, 0.9418\}$ and $\langle\hat{W}_{L-1}\rangle \in \{0.8647, -0.5202, 0.5520, -0.8965\}$.
The nontrivial observation is that summations of both $\langle\hat{W}_{1}\rangle$ and $\langle\hat{W}_{L-1}\rangle$ within the degenerate ground state
are extremely close to zero, which resemble those of zero edge modes.
Above the ground state, the lowest excitations come from flipping further $\mathbb{Z}_2$ quantities $\hat{W}_{3}$ and $\hat{W}_{L-3}$ near the boundaries.
To detect the trails of excitations, we define the local excitation energy $\omega_l$ as
\begin{eqnarray}\label{EQ:ExcEgEe}
\omega_{2l-1} &=& \langle S_{2l-1}^xS_{2l}^x\rangle_{e} - \langle S_{2l-1}^xS_{2l}^x\rangle_{g}, \nonumber\\
\omega_{2l}   &=& \langle S_{2l}^yS_{2l+1}^y\rangle_{e} - \langle S_{2l}^yS_{2l+1}^y\rangle_{g},
\end{eqnarray}
where $\langle \cdot \rangle_{g}$ and $\langle \cdot \rangle_{e}$ represent the expectation values with respect to the first and fifth energy levels, respectively.
Figure~\ref{FIG-KtvOBCGap}(a) shows the distribution of excitation energy $\omega_l$ for the isotropic Kitaev spin chain ($g = 1$) under OBC.
It can be found that the excitation energy is indeed localized at the very boundary parts, with $\omega_l$ being zero in the central region.
The total excitation gap $\Delta_e = \sum_l \omega_l$ is 0.11594909(2),
which is roughly $2/3$ of the bulk gap $0.17634970(2)$ reserved in the PBC case.

We propose that the loss of excitation gap is attributed to Ising-type couplings and short correlation length $\xi \simeq 1$ of the Kitaev spin chain.
Because of the OBC imposed, correlations within the marginal $x$-type bonds of $(1,2)$ and $(L-1,L)$ are enhanced,
leading to effective $x$-type couplings among these four sites.
Hence, the excitations are entangled and the marginal $\mathbb{Z}_2$ quantities are no longer conserved.
Because of the extremely short correlation length, the excitations are gathered at edges of the chain.
Unfortunately, the emergent $S_1^xS_L^x$ correlation does not contribute to the energy.
Instead, it trammels the excitations and thus reduce the excitation gap.
Our analysis implies that, if excitations are not bounded at the edges, then excitation gap $\Delta_e$ would not depend on the boundary condition severely.
To check the conjecture, we consider a stronger $y$-type bond, say $g = 1.2$.
We show the distribution of excitation energy $\omega_l$ in Fig.~\ref{FIG-KtvOBCGap}(b),
and it is clearly found that the excitation comes from the very middle of the chain.
More importantly, discrepancy of the excitation gap between the OBC and PBC is insignificant.
We have shown the excitation gap $\Delta_e$ as a function of anisotropy $g$ under both OBC (red circle) and PBC (blue square) in Fig.~\ref{FIG-KtvOBCGap}(c).
When $g \lesssim 1.06$, the excitations are localized at edges and the excitation gap under OBC is somewhat smaller.

\section{Effect of imbalanced bond strength}\label{SEC:AnisoKG}

\subsection{Haldane--Dimer transition}
In this section we study the effect of imbalanced bond strength with $g \neq 1$.
We recall that our model \eqref{J1J2KG-Ham} is reduced to the bond-alternating spin-1 Heisenberg chain when $\theta = -\pi/4$,
which undergoes a Haldane--dimer transition at $g = 0.587(2)$ that belongs to the Gaussian universality class
with a central charge $c = 1$ \cite{AffleckHaldane1987,KatoTanaka1994,KitazawaNO1996,SuChenXiang2016}.
The dimerized phase is characterized by a stable alternation of nearest-neighbor spin-spin correlations,
which is equivalent to the difference of $\langle\tilde{S}_i^z\tilde{S}_j^z\rangle$ between the odd bonds and even bonds within the six-site unit cell, namely
\begin{equation}\label{EQ:DimerOP}
O_D^{z} = \sum_{l=1,2,3} \big\vert\langle\tilde{S}_{2l-1}^z\tilde{S}_{2l}^z\rangle\big\vert - \big\vert\langle\tilde{S}_{2l}^z\tilde{S}_{2l+1}^z\rangle\big\vert.
\end{equation}
Due to the inherent bond alternation, the transitional symmetry of the neighboring sites is broken innately.
When $g < 1$, the $x$-type bond is stronger than the $y$-type, and vice versa.
Hence, the ground state of the dimerized phase is unique with a finite energy gap.

\begin{figure}[!ht]
\centering
\includegraphics[width=0.95\columnwidth, clip]{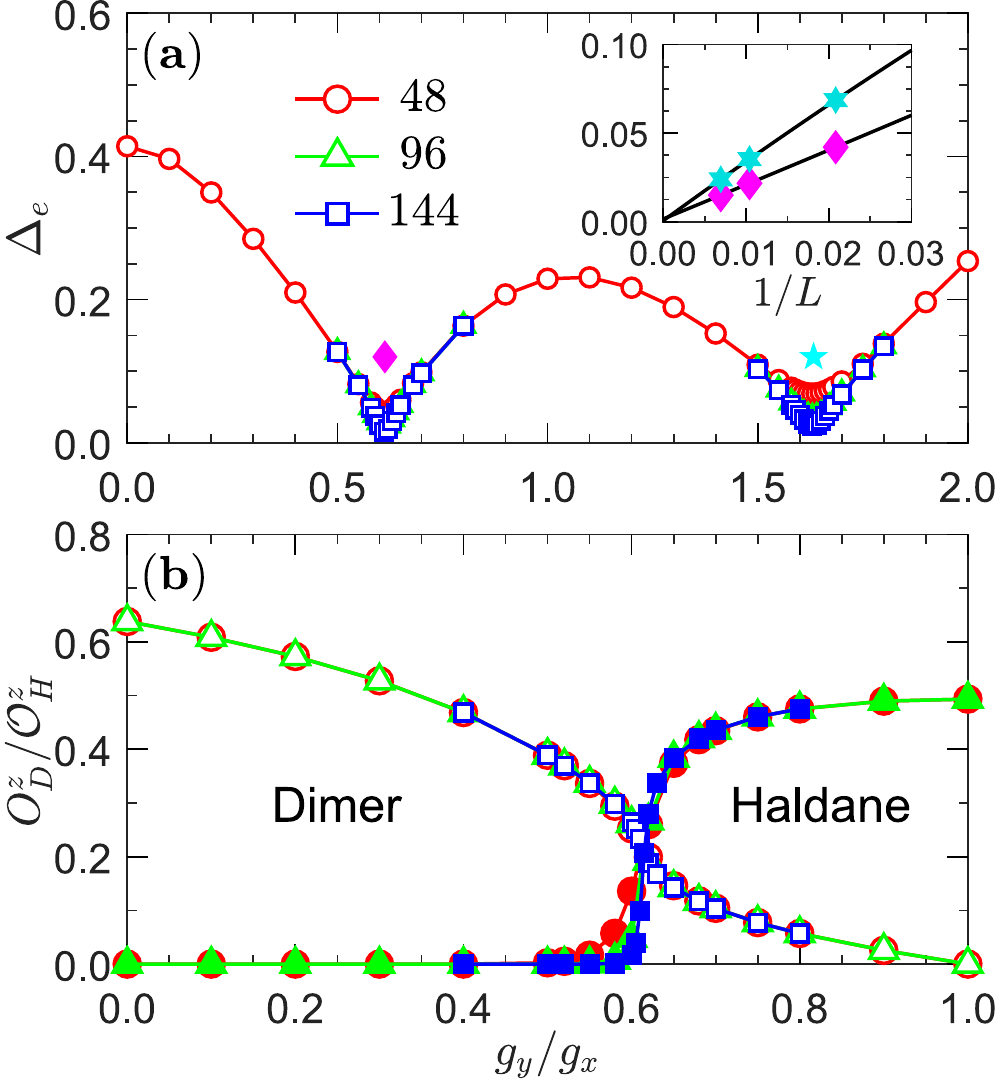}\\
\caption{(a) Excitation gap $\Delta_e$ of the dimerized phase and the Haldane phase in the pure $\Gamma$ chain with $\theta = 0$.
    The length $L$ of the chain is 48 (red circle), 96 (green triangle), and 144 (blue square).
    Inset: Extrapolations of the minimal excitation gaps around the transition points marked by the diamond and pentagram, respectively.
    (b) SOP $\mathcal{O}_H^{z}$ of the Haldane phase (filled symbols) and dimer order parameter $O_D^{z}$ of the dimerized phase (open symbols) as a function of $g$.}
    \label{FIG-HDvsDimer}
\end{figure}

Figure~\ref{FIG-HDvsDimer}(a) shows the energy gap $\Delta_e$ for three different length $L$ = 48, 96, and 144.
As $g$ goes from 0 to 2, $\Delta_e$ exhibits two remarkable valleys where the value becomes smaller and smaller as $L$ is increased.
We make a linear extrapolation of the series of minimal values around each valley marked by a pink diamond ($g < 1$) or a cyan pentagram ($g > 1$),
and it is shown in the inset that the excitation gap $\Delta_e$ indeed closes at the transition point in the thermodynamic limit.
Noteworthy, we find that the transition points are $g_{t,1} = 0.613(2)$ and $g_{t,2} = 1.633(2)$,
satisfying the self-dual relation as $g_{t,1}\cdot g_{t,2} \simeq 1$.
In Fig.~\ref{FIG-HDvsDimer}(b),
the SOP $\mathcal{O}_H^{z}$ (see Eq.~\eqref{EQ:SOPHaldane}, filled symbols) of the Haldane phase
and the dimer order parameter $O_D^{z}$ (see Eq.~\eqref{EQ:DimerOP}, open symbols) of the dimerized phase
are plotted in the window of $g\in[0, 1]$.
It is found that $\mathcal{O}_H^{z}$ is finite when $g > 0.613(2)$ and is vanishingly small otherwise.
Although $O_D^{z}$ is nonzero as long as $g \neq 1$, it changes abruptly on the brink of the Haldane phase.
We have taken the first-order derivative of $O_D^{z}$ (not shown), and the peak position coincides with
the critical point $g_{c} = 0.613(2)$ where the SOP of the Haldane phase nicely vanishes.

\subsection{Magnetically ordered phases}

As demonstrated, the dimerized phase is favored by sufficient bond alternation.
In the case of strong anisotropy where $g \lesssim 0.4$ or $g \gtrsim 2.5$,
there is a direct transition between the dimerized phase and the $\mathcal{M}_O$ phase.
Otherwise an intermediate region is stabilized as a consequence of the interplay between competing interactions and modest bond alternation.
A more careful inspection shows that it harbors two magnetically ordered states (see Fig.~\ref{FIG-GSPD}(b)) with distinct spin patterns.
On the side near the dimerized phase, it is a FM$_{U_6}$ phase whose spins align along one specific direction
of $\pm\hat{x}$, $\pm\hat{y}$, or $\pm\hat{z}$. For example, it could be
\begin{equation}\label{EQ:FMU6}
\big(\langle \tilde{\mathbf{S}}_1\rangle, \langle \tilde{\mathbf{S}}_2\rangle, \langle \tilde{\mathbf{S}}_3\rangle; \langle \tilde{\mathbf{S}}_4\rangle, \langle \tilde{\mathbf{S}}_5\rangle, \langle \tilde{\mathbf{S}}_6\rangle\big) \!=\! \big({c},{b},{b};{c},{a},{a}\big)\hat{z}
\end{equation}
where ${a}$, ${b}$ and ${c}$ are intensities of the magnetization along the $\hat{z}$ direction and $a, b, c \leq 1$.
The other is an in-plane magnetic phase (termed $\mathcal{M}_I$) where one of the spin components is zero.
We find that spins within the magnetic unit cell reads
\begin{eqnarray}\label{EQ:FMInPln123}
\langle \tilde{\mathbf{S}}_1\rangle \!=\!
    \left(
    \begin{array}{c}
        c \\
        0 \\
        c
    \end{array}
    \right),\quad
\langle \tilde{\mathbf{S}}_2\rangle \!=\!
    \left(
    \begin{array}{c}
        a \\
        0 \\
        b
    \end{array}
    \right),\quad
\langle \tilde{\mathbf{S}}_3\rangle \!=\!
    \left(
    \begin{array}{c}
        a \\
        0 \\
        b
    \end{array}
    \right)
\end{eqnarray}
and
\begin{eqnarray}\label{EQ:FMInPln456}
\langle \tilde{\mathbf{S}}_4\rangle \!=\!
    \left(
    \begin{array}{c}
        b \\
        0 \\
        a
    \end{array}
    \right),\quad
\langle \tilde{\mathbf{S}}_5\rangle \!=\!
    \left(
    \begin{array}{c}
        b \\
        0 \\
        a
    \end{array}
    \right),\quad
\langle \tilde{\mathbf{S}}_6\rangle \!=\!
    \left(
    \begin{array}{c}
        c \\
        0 \\
        c
    \end{array}
    \right).
\end{eqnarray}
For this phase, the value of $a, b, c$ should be bounded by $\sqrt{a^2+b^2} \leq 1$ and $c \leq 1/\sqrt2$.

\begin{figure}[!ht]
\centering
\includegraphics[width=0.95\columnwidth, clip]{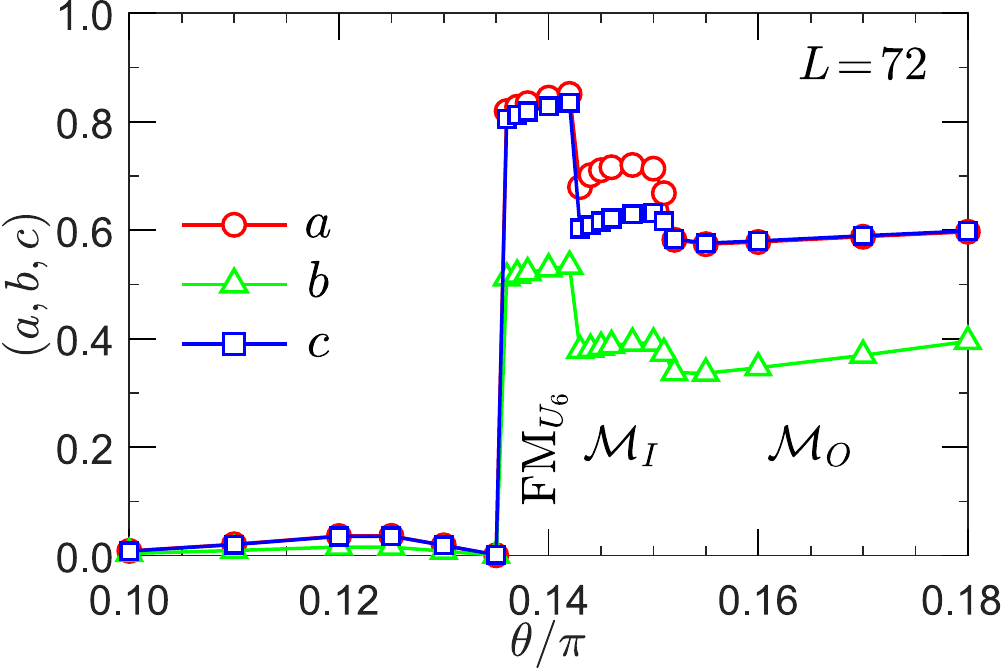}\\
\caption{Sublattice magnetization $(a, b, c)$ of the FM$_{U_6}$ phase, $\mathcal{M}_I$ phase, and $\mathcal{M}_O$ phase
    in a 72-site $K$-$\Gamma$ chain with $g = 2$.}
    \label{FIG-MagOrdABC}
\end{figure}

The series of QPTs between the magnetically ordered states could be signified by the magnetization of ($a$, $b$, $c$).
In this regard, we focus on the line of $g = 2.0$ (which is equivalent to $g = 0.5$ because of the self-dual relation)
and calculate these values from the spin-spin correlation functions on a chain of length $L = 72$.
Figure~\ref{FIG-MagOrdABC} shows the values of ($a$, $b$, $c$) as a function of $\theta$ in the range $0.10\pi \leq \theta \leq 0.18\pi$.
Due to the finite-size effect, $a$, $b$, and $c$ are nonzero but very tiny when $\theta/\pi < 0.1345(5)$.
We have checked that they will eventually go to zero in the thermodynamic limit.
As $\theta$ exceeds 0.1345(5), the system enters into the FM$_{U_6}$ phase whose values of ($a$, $b$, $c$) are accidentally bigger than $1/2$.
Afterwards, the system undergoes another two first-order QPTs at which the ground state evolves from the $\mathcal{M}_I$ phase to $\mathcal{M}_O$ phase, respectively.

\section{Double-peak specific heat in the Kitaev phase}\label{SEC:CvEE}

In this section we go beyond the ground-state study by calculating thermodynamic quantities in the spin-1 Kitaev chain
to elucidate the unusual excitations of the Kitaev phase.
We recall that specific heat $C_v$ of the spin-$1/2$ Kitaev honeycomb model is well-recognized
to exhibit a double-peak structure at two different energy scales,
signifying two kinds of Majorana fermions resulting from fractionalization of spin degrees of freedom \cite{KnolleKCM2014}.
The high-temperature peak relates to the enhancement of short-range spin-spin correlations
while the low-temperature peak comes from freezing of fluxes \cite{NasuUM2015}.
In addition, the thermal entropy displays an approximately half plateau with the value $\frac12\ln2 k_B$ per site in the intermediate crossover region,
in accordance with a half release of entropy around each peak.
Moreover, existence of double-peak specific heat is also reported in the spin-1 Kitaev honeycomb model,
yet precise position of the low-temperature peak is blurry because of the strong finite-size effect \cite{KogaTN2018}.
We note in passing that there has been a renascent interest in the thermodynamics of the Kitaev QSL quite recently
\cite{EschmannMB2019,LiQuZhang2020,FengPerBur2020,JahromiYO2020}.

In contrast to the spin-$1/2$ Kitaev chain which only possesses a sole peak in the specific heat (see Appendix \ref{AppA} for detail),
our study unambiguously suggests that the spin-1 Kitaev chain displays a double-peak specific heat.
To illustrate this, we begin by introducing the partition function
$\Xi = \textrm{Tr} e^{-\beta \mathcal{H}}$ with $\beta = 1/k_BT$ (herein, the Boltzmann constant $k_B$ = 1),
which is generally the starting point to calculate thermodynamic quantities.
Consequently, the free energy is $F = -\beta^{-1}\ln\Xi$ and the internal energy is
$U = -\frac{\partial \ln \Xi}{\partial \beta}$.
The specific heat is thus calculated by
\begin{align}
C_v = \frac{1}{N}\left(\frac{\partial U}{\partial T}\right)_V = -\frac{\beta^2}{N} \frac{\partial U}{\partial \beta},
\end{align}
and the thermal entropy is given by
\begin{align}\label{EQ:Entropy}
\mathcal{S} = \frac{\beta}{N}(U - F) = \mathcal{S}_0 + \int_0^{T} \frac{C_v(T')}{T'} dT',
\end{align}
with $\mathcal{S}_0$ being the residual entropy at zero temperature.

The direct way to calculate the partition function $\Xi$ is by diagonalizing the Hamiltonian
from which the entire energy spectrum $\{E_{\upsilon}\}$ is readily available.
Needless to say, this route is strongly limited by the system size and we consider a six-site closed chain for simplicity.
Surprisingly, physical quantities such as $C_v$ and $\mathcal{S}$ suffer from a weak finite-size effect
and the results are fairly close to those in the thermodynamic limit.
The first few energy levels are shown in Tab.~\ref{Tab-KtvSixSite},
and their degeneracies $\rho_{\upsilon}$ are $\{1, 6, 6, 2, \cdots\}$.

\begin{table}[th!]
\caption{\label{Tab-KtvSixSite}
Energy spectrum of the spin-1 Kitaev spin model on a six-site closed chain.
The first four columns are energy level index $\upsilon$, energy $E_{\upsilon}$, degeneracy $\rho_{\upsilon}$, and energy gap $\Delta_{\upsilon} = E_{\upsilon} - E_0$.
The last column ($\tilde{\Delta}_{\upsilon}$) represents the approximation of the energy gap $\Delta_{\upsilon}$ in a unit of $\Delta_{\kappa} \approx 0.18018574$.}
\begin{ruledtabular}
\begin{tabular}{ c c c  c  c}
$\upsilon$  & $E_{\upsilon}$    & $\rho_{\upsilon}$         & $\Delta_{\upsilon}$   & $\tilde{\Delta}_{\upsilon}/\Delta_{\kappa}$\\
\colrule
0           & -3.63027662       & 1                         & 0.00000000            &          $0$      \\
1           & -3.45009088       & 6                         & 0.18018574            &          $1$      \\
2           & -3.38928222       & 6                         & 0.24099440            & $\sim$ $4/3$      \\
3           & -3.33005874       & 2                         & 0.30021788            & $\sim$ $5/3$      \\
4           & $\cdots$          & $\cdots$                  & $\cdots$              & $\cdots$          \\
\end{tabular}
\end{ruledtabular}
\end{table}

A more reliable way to obtain $\Xi$ is by virtue of finite-temperature computational techniques such as the TMRG method \cite{BurXiangGeh1996,WangXiang1997},
which is an extension of the DMRG method to finite temperature ($T \neq 0$).
The TMRG method relies on the quantum-classical correspondence by way of the Trotter-Suzuki decomposition,
and represents the partition function as a trace of a series of transfer matrices.
It deals directly with an infinite spin chain and the errors come from
the Trotter-Suzuki step $\tau = \beta/M$ ($M$ is the Trotter-Suzuki number) and truncated number of states $m$ \cite{BurXiangGeh1996,WangXiang1997}.
We note that an additional reorthogonalization procedure is applied so that more block states can be kept \cite{HoneckerHPR2011}.
The TMRG method has been successfully applied to various quantum spin chains where several thermodynamic quantities such as specific heat and magnetic susceptibility
could be calculated with high precision \cite{Xiang1998,Johnston2000,GuSuGao2006,XiHuLZW2017}.
For the sake of accuracy, we fix $\tau = 0.01$ and $m = 1024$, which is enough to give a satisfactory precision
in our simulation down to the lowest temperature of $0.0033$.

\begin{figure}[!ht]
\centering
\includegraphics[width=0.95\columnwidth, clip]{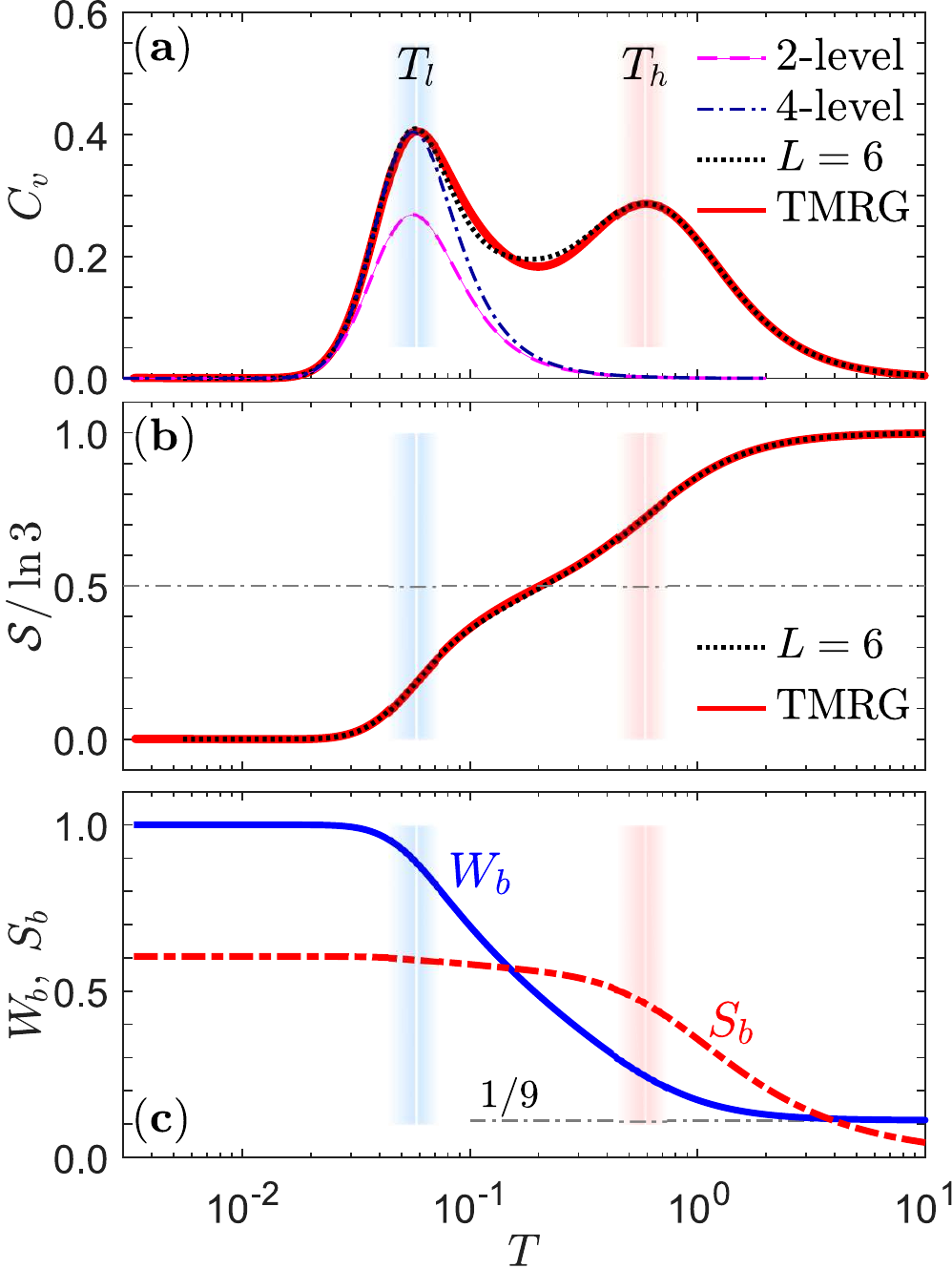}\\
\caption{Temperature dependencies of (a) specific heat $C_v$, (b) thermal entropy $\mathcal{S}$,
    and (c) expectation values of $\overline{W}_b$ and $S_b$ obtained by TMRG method.
    Results on a six-site closed chain (dotted line, black) are also shown in panel (a) and (b) for comparison.
    Based on the energy spectrum shown in Tab.~\ref{Tab-KtvSixSite},
    panel (a) also presents the specific heat of two-level (pink dashed) and four-level (blue dash-dotted) systems around the low-$T$ peak.}
    \label{FIG-KtvCvEntWp}
\end{figure}

Figure.~\ref{FIG-KtvCvEntWp}(a) shows the specific heat $C_v$ as a function of temperature $T$ up to 10.
Here, exact result on an extremely small system size of $L = 6$ (black dotted line)
and the TMRG calculation on an infinite size system (red solid line) are shown.
It can be found that $C_v$ is almost system-size independent as the difference between the two limit cases is quite small.
The specific heat $C_v$ acquires two peaks at a low temperature $T_l \simeq 0.0582$ and a high temperature $T_h \simeq 0.5860$, respectively.
The low-temperature peak is more pronounced and we propose that it relates to the large degeneracy of the low-lying excited states.
To demonstrate this, we firstly consider a two-level system which consists of the ground state and the first excited state
out of the six-site closed chain (see Tab.~\ref{Tab-KtvSixSite}).
The degeneracies of the two states are $\rho_0 = 1$ and $\rho_1 = 6$, respectively, with an energy gap $\Delta_{\kappa} \approx 0.18018574$ between them.
For this system, the partition function $\Xi = \rho_0 + \rho_1 e^{-\beta\Delta_{\kappa}}$, and the specific heat is
\begin{align}\label{EQ:CvTwoLevel}
C_v = \frac{\rho_1}{\rho_0} \frac{(\beta\Delta_{\kappa})^2}{\Big(e^{\beta\Delta_{\kappa}/2} + \frac{\rho_1}{\rho_0} e^{-\beta\Delta_{\kappa}/2} \Big)^2},
\end{align}
which depends on the energy gap $\Delta_{\kappa}$ and relative degeneracy $\rho_1/\rho_0 = 6$.
The specific heat in Eq.~\eqref{EQ:CvTwoLevel} shows a peak at the extreme temperature $T_p = \Delta_{\kappa}/x_p$,
where $x_p = 3.23565205\cdots$ is determined by the transcendental equation
$\frac{x-2}{x+2}e^{x} = {\rho_1}/{\rho_0} = 6$.
This yields the extreme temperature $T_p \approx 0.0557$, which is fairly close to $T_l \simeq 0.0582$.
The drawback of this oversimplified approximation is that intensity of the specific heat is smaller than the actual value.
However, the intensity could be significantly improved by considering a four-level system with a partition function
\begin{align}
\Xi = \rho_0 + \rho_1 e^{-\beta\Delta_{\kappa}} + \rho_2 e^{-4\beta\Delta_{\kappa}/3} + \rho_3 e^{-5\beta\Delta_{\kappa}/3}.
\end{align}
The corresponding specific heat is also shown in Fig.~\ref{FIG-KtvCvEntWp}(a) (blue dash-dotted line),
and position and extremum are both close to the TMRG result on the infinite system.
With increasing system size, the dimension of the Hilbert space will enlarge exponentially,
and a further multi-level system should be considered to reproduce the low-$T$ peak
(see Appendix \ref{AppB} for a discussion on a 12-site closed chain).

\begin{figure}[!ht]
\centering
\includegraphics[width=0.95\columnwidth, clip]{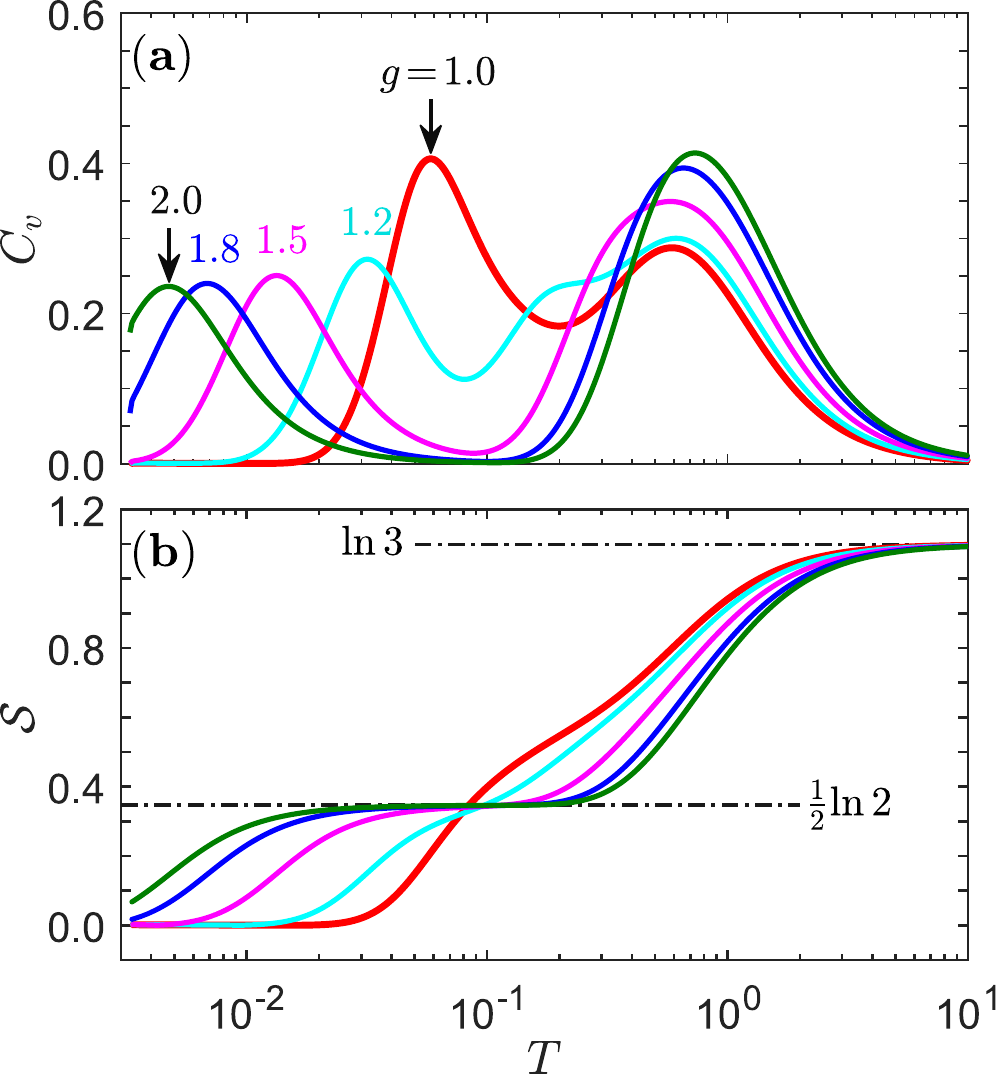}\\
\caption{Temperature dependencies of (a) specific heat $C_v$ and (b) entropy $\mathcal{S}$ of the anisotropic spin-1 Kitaev chain in the infinite-size system.
    The anisotropy $g$ is 1.0 (red), 1.2 (cyan), 1.5 (pink), 1.8 (blue), and 2.0 (green).}
    \label{FIG-AniKtvCvEnt}
\end{figure}

Figure.~\ref{FIG-KtvCvEntWp}(b) shows the behavior of entropy $\mathcal{S}$ defined in Eq.~\eqref{EQ:Entropy}.
The entropy decreases from its saturated value of $\ln3$ rapidly with the lowering of the temperature in the neighborhood of $T_h$ and $T_l$.
The $1/2$-plateau of the entropy is smeared out in the intermediate region between the two temperature scales.
Instead, a shoulder in entropy is observed and the entropy becomes $\sim\frac12\ln3$ at a temperature of $T_m \simeq 0.20$.
Interestingly, it is exactly the same temperature at which specific heat shows its local minimum.
To understand physical mechanisms of the double-peak structure in the specific heat,
we calculate the averaged bond density $\overline{W}_b$ (see Eq.~\eqref{EQ:WbAvg}) and the averaged nearest-neighbor correlator
\begin{align}\label{EQ:CorrFuncBond}
S_b = \frac{1}{L}\sum_{\langle ij\rangle_{\gamma}} \langle S_i^{\gamma}S_j^{\gamma} \rangle.
\end{align}
The results of $\overline{W}_b$ and $S_b$ are shown in Fig.~\ref{FIG-KtvCvEntWp}(c).
Below the low-$T$ peak at $T_l$, $\overline{W}_b$ is almost unchanged with a value of $\sim\!1$.
It then decreases to $1/9$ successively with the increase in temperature in the intermediate region\footnote{
In the high-$T$ paramagnetic phase, we have $(S_l^x)^2$ = $(S_l^y)^2$ = $(S_l^z)^2$ = $2/3$ for $\forall~l$ in the spin-1 system.
The bond-parity operator $\hat{W}_l$ is also uniformly distributed.
Taking the $x$-type operator for instance, $\hat{W}_l$ = $(\Sigma_l^x)^2$ = $[1-2(S_l^x)^2]^2$ = $1/9$.}.
Therefore, the low-$T$ peak originates from the local conserved quantity.
On the other hand, the high-$T$ peak is closely related to the growth of short-range spin-spin correlations.
Above the high-$T$ peak at $T_h$, $S_b$ is very small, signifying a paramagnetic phase.
It is then enhanced dramatically by decreasing the temperature and is stabilized around 0.60,
which is merely the absolute value of the ground-state energy of the isotropic spin-1 Kitaev chain.

We would emphasize that the double-peak specific heat is a universal behavior of the Kitaev phase in spite of anisotropy $g$.
For this purpose, we have calculated the specific heat and entropy for several different $g$ up to 2.0, see Fig.~\ref{FIG-AniKtvCvEnt}.
It can be found that the high-$T$ peak is very pronounced while the low-$T$ peak is shifted to a lower temperature as $g$ is increased.
As can be seen from Fig.~\ref{FIG-AniKtvCvEnt}(b), there is a $\frac12\ln2$-plateau of entropy in the middle region,
and the plateau will last for a larger temperature scale with increasing anisotropy.
Thus, in the large anisotropy limit where $g$ is infinity (or equivalently, $g \to 0$),
the low-$T$ peak vanishes and there is a residual entropy at the zero temperature due to the $2^{L/2}$-fold ground-state degeneracy.
In this circumstance, a spin-1 Ising bond of either $x$-type or $y$-type is capable of revealing the residual entropy.
The energy spectrum of this two-site model is $\{-1, 0, 1\}$, with a degeneracy of $\{2, 5, 2\}$, respectively.
Thus, the partition function is known to be $\Xi = 4\cosh(\beta K) +5$.
According to Eq.~\eqref{EQ:Entropy}, we arrive at the entropy
\begin{equation}\label{EQ:EntKtvBond}
\mathcal{S} = \frac12\ln\big(4\cosh(\beta K) +5\big) - \frac{2(\beta K)\cosh(\beta K)}{4\cosh(\beta K) +5}.
\end{equation}
For large enough temperature, the entropy in Eq.~\eqref{EQ:EntKtvBond} is expected to yield $\ln 3$.
As $T$ becomes infinitely small, we find that $\mathcal{S}(T\to0) = \frac12\ln2$.
In this circumstance, the low-temperature peak in the specific heat disappears,
and the sole peak locates at $T\simeq 0.3896$ with the maximal value $\bar{C}_v \simeq 0.4974$.

\section{Conclusion}\label{SEC:CONC}

In this paper, we studied a bond-alternating spin-1 $K$-$\Gamma$ chain,
focusing on the nonmagnetic Haldane phase and Kitaev phase that exhibit unusual excitations.
The Haldane phase is an outstanding example of a SPT phase which is gapped with short-range entanglement,
while the Kitaev phase is a one-dimensional incarnation of the Kitaev QSL and is not a SPT phase since its lowest entanglement spectrum is unique.
Whereas both of the phases have unique ground states under the PBC, the degeneracies of the first-excited states are different.
It is triplet degenerate for the former, while it is $L$-fold degenerate with $L$ being the chain length for the latter as a result of the bond-resolved $\mathbb{Z}_2$ quantities.
Interestingly, they both possess four-fold degenerate ground states in the case of the OBC.
The degeneracy in the Haldane phase comes from the spin-$1/2$ edge states,
while it originates from two marginal $\mathbb{Z}_2$ quantities in the Kitaev phase.
On top of the degenerate ground state in the Kitaev phase,
the spatial profile of the excitations highly relies on the relative bond strength $g \equiv g_y/g_x$.
The excitations are confined at the boundaries of the chain when $g \lesssim 1.06$ and locate at the very middle otherwise.
In the former case, the excitations at edges are entangled, weakening the excitation gap as compared with its PBC counterpart.
The quantum phase diagram also contains two dimerized phases
which undergo continuous QPTs to the Haldane phase when tuning the anisotropy $g$.
In addition, three magnetically ordered states are identified,
of which the $\mathcal{M}_O$ phase is the most energetically favored and is situated alongside the AFM Kitaev phase.

We also investigated the thermodynamic behaviors of the Kitaev phase,
which is found to exhibit a fascinating double-peak structure in the specific heat.
During the low-temperature and high-temperature crossover region, the entropy is released gradually without generating a plateau.
Pertaining to the origin of the double peaks,
we propose that the high-temperature peak relates to the enhancement of nearest-neighbor spin correlation
while the low-temperature peak comes from freezing of $\mathbb{Z}_2$ quantities
and is relevant to the highly degenerate low-lying excited states.
We also find that the double-peak specific heat is robust against anisotropy $g$,
although the low-temperature peak is shifted to lower temperature steadily as $g$ deviates from 1.
Notably, a $\frac12\ln2$-plateau of entropy appears in the crossover region.

In closing, we would like to make some remarks on the connection between the Kitaev phase and the sought-after QSL
in the spin-1 Kitaev honeycomb model \cite{KogaTN2018,ZhuWengSheng2020,HickeyTrebst2020,KhaitKim2021}.
The spin-1 Kitaev QSL has been gaining much attention over the years because of the growing interest in high-spin Kitaev materials \cite{StaPerKee2019}.
For both phases, there is no spontaneously symmetry breaking in the ground states, and they are gapped with extremely short-range spin-spin correlations.
Noteworthily, they both exhibit anomalous double peaks in their specific heat.
However, the intrinsic nature of the one-dimensional gapped Kitaev phase remains to be explored in future works.
For example, it would be intriguing to know to which class the Kitaev phase belongs from the perspective of symmetry fractionalization \cite{ChenGuWen2011}.
Also, calculation of the low-energy excitation spectrum from the dynamic structure factor is capable of probing elementary excitations \cite{Punk2014}.
On all counts, by using a high-precision numerical study of the Kitaev phase,
our work would offer some insights into the spin-1 Kitaev QSL on the honeycomb lattice.

\begin{acknowledgments}
Q.L. would like to thank Z.-X. Liu and W.-L. You for useful discussions,
and is also indebted to X. Wang and J. Zhao for an inspiring discussion
which helped initiate this work.
S.H. acknowledged support from Grant No. NSAF-U1930402.
H.-Y.K. was supported by NSERC Discovery Grant No. 06089-2016,
the Centre for Quantum Materials at the University of Toronto,
the Canadian Institute for Advanced Research,
and also funding from the Canada Research Chairs Program.
The computations were partly performed on the Tianhe-2JK at the Beijing Computational Science Research Center (CSRC).
Computations were also performed on the Niagara supercomputer at the SciNet HPC Consortium.
SciNet is funded by: the Canada Foundation for Innovation under the auspices of Compute Canada;
the Government of Ontario; Ontario Research Fund - Research Excellence; and the University of Toronto.
\end{acknowledgments}

\begin{figure}[!ht]
\centering
\includegraphics[width=0.95\columnwidth, clip]{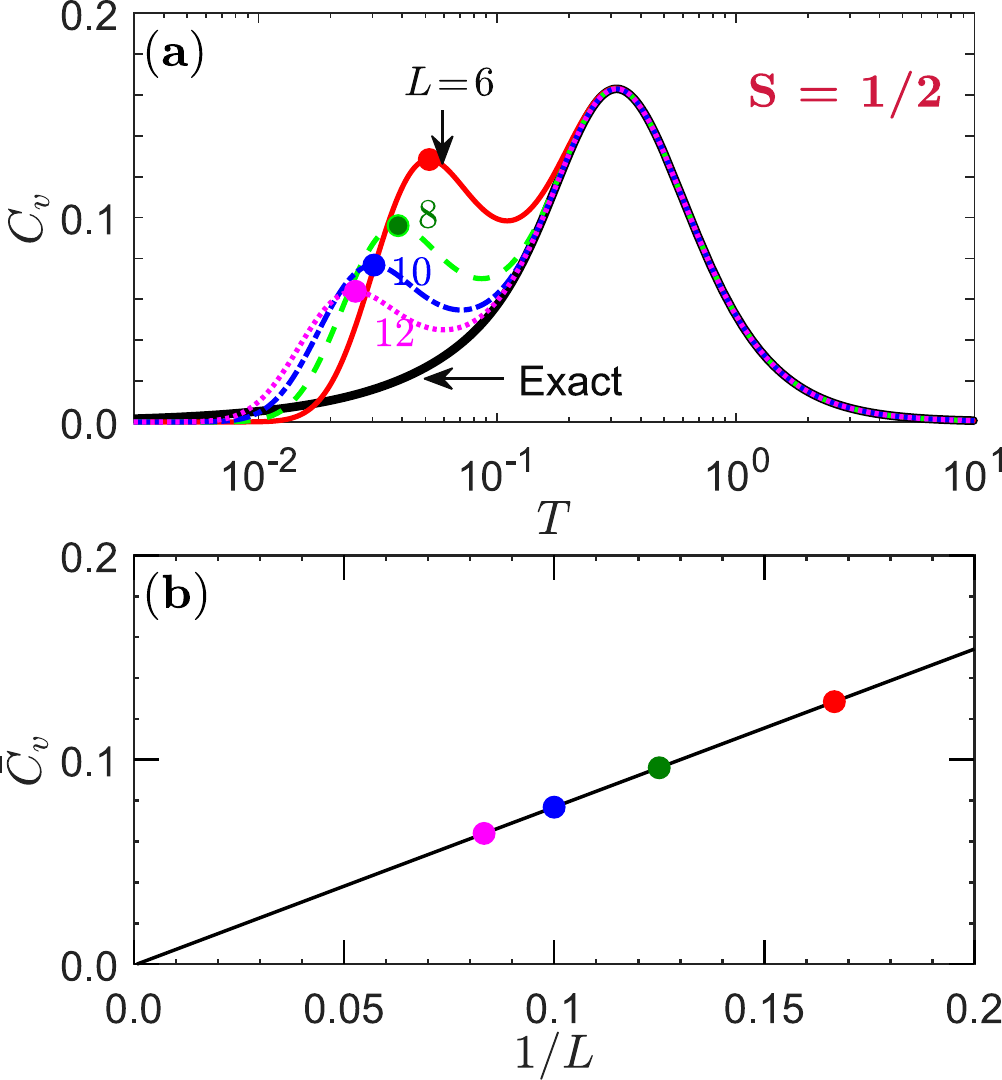}\\
\caption{(a) Specific heat $C_v$ of the isotropic ($g = 1$) spin-$1/2$ Kitaev chain on a closed chain of length $L$ = 6 (red), 8 (green), 10 (blue), and 12 (pink).
  The filled circles mark the extreme subleading peaks $\bar{C}_v$ at given chain length.
  The thick solid line represents the exact specific heat (see Eq.~\eqref{EQA:CvTFIM}) in the thermodynamic limit.
  (b) Linear extrapolation of the subleading peaks $\bar{C}_v$ to the infinite-size limit.}
  \label{FIG-Spin12Cv}
\end{figure}


\appendix
\setcounter{equation}{0}
\renewcommand{\theequation}{A\arabic{equation}}

\section{Specific heat of spin-$1/2$ Kitaev chain}\label{AppA}

\begin{figure}[!ht]
\centering
\includegraphics[width=0.95\columnwidth, clip]{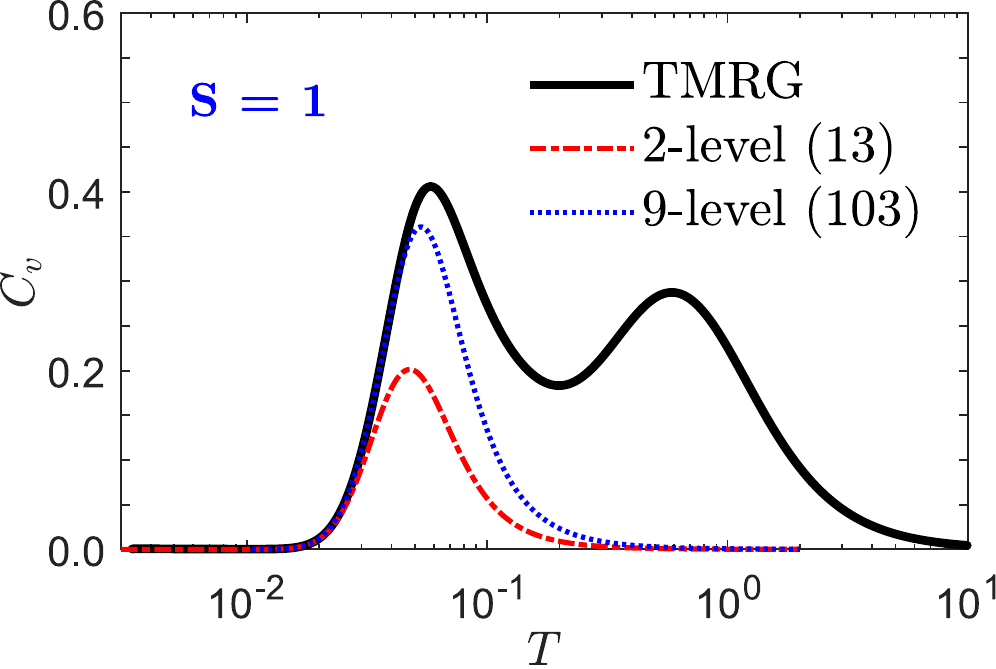}\\
\caption{Specific heat $C_v$ of the isotropic ($g = 1$) spin-1 Kitaev chain.
  The thick black line represents the TMRG result on an infinite-size system.
  Based on the energy spectrum of a 12-site closed chain,
  the specific heat on a two-level ansatz (with the first 13 energy levels, red dot-dashed line)
  and a nine-level ansatz (with the first 103 energy levels, blue dotted line) is shown around the low-temperature peak.}
  \label{FIG-LowTL12}
\end{figure}

The Hamiltonian of the Kitaev spin chain reads \cite{BrzezickiDO2007,YouTian2008}
\begin{equation}\label{EQA:KtvSpinChn}
\mathcal{H}_K = \sum_{l=1}^{L/2} \big( g_x S_{2l-1}^x S_{2l}^x + g_y S_{2l}^y S_{2l+1}^y \big)
\end{equation}
where $\textbf{S}_l = (S_l^x, S_l^y, S_l^z)$ is the spin-$1/2$ operator at site $l$, and $L$ is the total number of sites.
By using a spin duality transformation, it could be rewritten as \cite{FengZX2007}
\begin{equation}\label{EQA:KtvSpinChn}
\mathcal{H}_K = \sum_{l=1}^{L} \left( g_x \tilde{S}_{2l}^x \tilde{S}_{2l+2}^x + \frac{g_y}{2} \tilde{S}_{2l}^y \right),
\end{equation}
which is a \textit{diluted} transverse field Ising model.
For this model, the specific heat is exactly known as \cite{Pfeuty1970}
\begin{equation}\label{EQA:CvTFIM}
C_v = \frac{1}{2}\int_0^{\pi}\frac{dk}{\pi} \Big(\frac{\beta\epsilon_k}{2}\Big)^2 \textrm{sech}^2 \Big(\frac{\beta\epsilon_k}{2}\Big)
\end{equation}
where $\beta = 1/(k_BT)$, and the dispersion energy is $\epsilon_k = g_x\sqrt{1+g^2+2g\cos k}/2$
with $g = g_y/g_x$.
We note that the prefactor $1/2$ in Eq.~\eqref{EQA:CvTFIM} comes from the fact that only one half of the spins (i.e., spins at even sites)
are involved in the Hamiltonian of Eq.~\eqref{EQA:KtvSpinChn}.

Figure \ref{FIG-Spin12Cv}(a) shows the specific heat $C_v$ of the isotropic ($g = 1$) spin-$1/2$ Kitaev chain at finite-size system
of $L = 6$ (red), $L = 8$ (green), $L = 10$ (blue), and $L = 12$ (pink).
It can be found that there is a pronounced peak at $T \simeq 0.3162$,
below which there is a subleading peak at a lower temperature which becomes smaller and smaller as $L$ is increased.
We emphasize that the low-temperature peak is a finite-size effect and it will disappear as $L$ goes to infinity (see the thick black line).
As shown in Fig.~\ref{FIG-Spin12Cv}(b), a linear extrapolation of the subleading peaks indeed gives a zero value when $L\to\infty$.
In the low temperature region where $T \lesssim 0.1$,
the specific heat is subject to the asymptotic behavior $C_v(T) \simeq \pi T/6$ \cite{KoppChak2005},
signifying a gapless system.
This, in turn, demonstrates the dramatic difference between the spin-$1/2$ and spin-1 Kitaev chains
as the latter is gapped and presents a stable double-peak structure in the specific heat.

\section{Low-temperature peak in a 12-site spin-1 Kitaev chain}\label{AppB}

In Fig.~\ref{FIG-KtvCvEntWp}(a), we have demonstrated that the low-$T$ peak of the specific heat in the spin-1 Kitaev chain
relates to the large degeneracy of the low-lying excited states.
To further strengthen such a conclusion, we now show the evolution of the low-$T$ peak by increasing the number of energy levels in a 12-site closed chain.
For this system, the ground state is unique while the first-excited state is 12-fold degenerate,
separated by an excitation gap $\Delta_{\kappa} \approx 0.17630343$.
Furthermore, the degeneracies of the first nine energy levels are successively $\{1, 12, 12, 12, 6, 12, 12, 24, 12\}$.
Figure~\ref{FIG-LowTL12} shows the low-$T$ peak based on a two-level ansatz (with the first 13 energy levels, red dot-dashed line)
and a nine-level ansatz (with the first 103 energy levels, blue dotted line).
The TMRG result (thick black line) is also shown for comparison.
We find that the two-level system could roughly recover the low-$T$ peak, although the position and height of the peak deviate from the TMRG result.
However, the nine-level system with 103 energy levels could significantly improve the result.
We note that the number of energy levels is only a rather small portion ($\sim2\times10^{-4}$ )
of the whole energy spectrum whose dimension is $3^{12} = 531441$.
With the increase in the system size, we believe that an even smaller portion of the whole energy levels will nicely produce the low-$T$ peak.
In this sense, we highlight the importance of the degenerate low-lying excited states in generating the low-temperature peak of the specific heat.




%

\end{document}